# Lithium-ion battery performance model including solvent segregation effects

Ruihe Li[a,c] [†], Simon O'Kane[a,c], Andrew Wang[b,c], Taeho Jung[b,c], Niall Kirkaldy[a,c], Monica Marinescu[a,c], Charles W. Monroe[b,c], Gregory J. Offer[a,c] [‡]

[a] Department of Mechanical Engineering, Imperial College London, UK

[b] Department of Engineering Science, University of Oxford, UK

[c] The Faraday Institution, UK

[†] r.li20@imperial.ac.uk, [‡] gregory.offer@imperial.ac.uk

## Abstract

A model of a lithium-ion battery containing a cosolvent electrolyte is developed and implemented within the open-source PyBaMM platform. Lithium-ion electrolytes are essential to battery operation and normally contain at least two solvents to satisfy performance requirements. The widely used Doyle–Fuller–Newman battery model assumes that the electrolyte comprises a salt dissolved in a single effective solvent, however. This single-solvent approximation has been disproved experimentally and may hinder accurate battery modelling. Here, we present a two-solvent model that resolves the transport of ethylene carbonate (EC) and lithium salt in a background linear carbonate. EC concentration polarization opposes that of $Li^+$ during cycling, affecting local electrolyte properties and cell-level overpotentials. Concentration gradients of $Li^+$ can be affected by cross-diffusion, whereby EC gradients enhance or impede salt flux. A rationally parametrized model that includes EC transport predicts 6% more power loss at 4.5C discharge and ~0.32% more capacity loss after a thousand 1C cycles than its single-solvent equivalent. This work provides a tool to model more transport behaviour in the electrolyte that may affect degradation and enables the transfer of microscopic knowledge about solvation structure-dependent performance to the macroscale.

# Introduction

Commercial lithium-ion batteries always combine multiple solvent components in the electrolyte to balance operational requirements and tailor for different electrode materials. For example, ethylene carbonate (EC) raises conductivity, ethyl-methyl carbonate (EMC) reduces viscosity and fluoroethylene carbonate (FEC) improves interfacial stability[1]. In practical battery electrolytes, components like these can be present with comparable molarities.

The Doyle-Fuller-Newman (DFN) framework, also called the pseudo-2D model or dualfoil, is the dominant physics-based model used to describe lithium-ion batteries at the cell level. Electrolyte-phase ion transport in the DFN framework is described using a version of concentrated-solution theory that assumes the electrolyte to comprise one solvent and one salt[2]. This 'single-solvent' approximation requires that all solvent components move as a single entity, i.e., that the molar ratio of cosolvents remains constant throughout a Li-ion cell.

Experimental and modelling studies have invalidated the single-solvent assumption, both for long-used commercial electrolytes and newly emerging electrolyte formulations. Various experimental techniques, including nuclear magnetic resonance (NMR)[3] and electrospray ionization mass spectroscopy (ESI-MS)[4], have shown that the composition of the $Li^+$ solvation sheath – the layer of molecules that surrounds $Li^+$ cations in solution, allowing them to dissolve – significantly differs from the bulk. The polar, cyclic carbonate components such as EC preferentially solvate $Li^+$ relative to nonpolar, linear counterparts like DMC, EMC, and DEC[5]. Preferential solvation by cyclic carbonates is commonly understood to originate from their higher permittivity[6], and has also been identified in the sodium-ion electrolyte $NaPF_6$ in 1:1(wt) EC:PC [7]. Apart from these microscopic, equilibrium observations, dynamical effects can arise from cosolvent interactions. Wang et al.[8] applied current pulses to $LiPF_6$:EC:EMC blends in a Hittorf cell and found that the pulse could induce a change in the EC:EMC ratio of as much as 50% at the electrode surface relative to the bulk.

Recent experimental observations make it imperative to rethink the single-solvent approximation that underpins the DFN framework, especially because the Li$^+$ solvation environment near electrode surfaces can substantially impact lithium-ion-battery performance[9]. Solvation effects have been closely studied at equilibrium on the microscale[9], but not yet dynamically at the cell level. When lithium intercalation occurs, the molecules that solvate electrolyte-phase lithium cations are released, and lithium deintercalation is accompanied by solvation events. Dynamical buildup or depletion of solvate components can impact interfacial side reactions, such as those that form the solid-electrolyte interphase (SEI)[5] and cathode-electrolyte interphase (CEI)[10]. Solvent composition also affects macroscopic transport properties. Experimental measurements have shown that the ionic conductivity of lithium-ion electrolytes is highly sensitive to cosolvent composition, especially when EC is added to linear carbonates[10-13].

Hayamizu[12] showed for EC:DEC blends that as the cosolvent ratio changed from 0:1 to 0.8:0.2 (molar basis), the diffusivity of Li$^+$ decreased by more than 50%, whereas the conductivity increased from 0.31 Sm$^{-1}$ to 1.0 Sm$^{-1}$. Wang et al.[8] found that an increase in EC:EMC molar ratio from 0:1 to 0.7:0.3 caused the conductivity of cosolvent ~1M LiPF$_6$ solutions to increase almost twofold (from 0.47 Sm$^{-1}$ to 0.92 Sm$^{-1}$). Such dramatic effects have been ignored in DFN models, leading to a research gap. It would be useful to have a cell level model that (1) evaluates the performance impacts of the individual solvents or additives that make up a cosolvent and (2) enables translation of knowledge about solvation structure from the microscale to the macroscale.

Fortunately, multi-component electrolytic transport has been well described within the Onsager−Stefan−Maxwell (OSM) framework, which is the foundation of Newman's concentrated-solution theory and the later DFN model[2]. As Wang et al.[8] pointed out, "[the single-solvent] approximation is born out of prohibitive experimental complexity rather than decisive experimental justification." Even in the ostensibly simple case of isothermal, isobaric, locally electroneutral transport in a single-solvent electrolyte, complete parametrization of the OSM flux laws requires three independent transport coefficients and three thermodynamic parameters – all of which may depend on composition. Measuring the composition dependence of these six parameters requires at least six carefully designed independent experiments. Such measurements and parameter extraction programmes have been implemented by Ma et al. [14], Valøen and Reimers [15], Nyman et al. [16], Monroe and co-authors [17-19]. Adding just one more solvent component doubles the number of binary species-species interactions at the molecular level, and consequently doubles the total requisite number of parameters. Huge additional effort would be needed to establish composition and temperature dependency of this parameter set.

Most of the very few papers that address multi-component transport within lithium-ion batteries have dealt with the lack of information about parameters by assuming that additives are dilute[20-22], i.e., by ignoring interactions between extra neutral components and ions in the mass conservation equation. When writing current-voltage relations for multicomponent electrolytes, these studies – presumably due to a lack of experimental data – have only accounted for concentration overpotential arising from concentration polarization of the dissolved lithium salt, neglecting the possibility that polarization of the cosolvent ratio can produce overpotential as well.

Two recent papers have clarified how cosolvent interactions can be integrated into electrolytic transport models. Wang et al.[8] studied $EMC:EC:LiPF_6$ ternary solutions with Hittorf cells, quantifying a cation transference number and an electro-osmotic coefficient, which respectively describe how applied current drives $Li^+$ and EC to migrate through the EMC background solvent. For the same ternary system, Jung et al.[23] measured the liquid-junction potential as a function of both EC and $Li^+$ concentration. These data can either be used to generate thermodynamic factors or incorporated directly into the current-voltage relation. Taken together, these property sets allow a fairly comprehensive description of transport within a co-solvent electrolyte (two solvents and one salt) within the DFN framework.

Here we extend the DFN model by incorporating the cosolvent transport model of Jung et al.[23] to describe the electrolytic phase. After presenting the mathematical description of this cosolvent DFN model, which we validate with experimental data, variation in the EC concentration is shown to affect internal states and observable performance characteristics of lithium-ion batteries, such as rate performance and ageing rates.

## Model description

To model the transport of four species (two types of solvents, one type of anion, and one type of cation.), we adapt the theoretical framework proposed by Monroe[24]. Compared with the widely used DFN model[25], which includes a cation balance, the Monroe framework requires an additional mass conservation equation for one of the solvents (in our case, EC). The flux laws that describe diffusion and migration of lithium and EC are coupled together by cross-terms that account for solute-solute interactions. Below we will describe the electrolyte model in detail; the additional relationships that describe the solid phases and intercalation processes match the typical DFN equations and are presented in the supplementary material.

SEI formation kinetics is also expected to depend strongly on the solvent composition near the interface. The model we present here also includes the interstitial diffusion limited SEI reaction proposed by von Kolzenberg et al.[26], in which we have incorporated a linear dependence of rate constants on the surface EC concentration.

**Electrolyte-phase transport model**

Following the DFN approach, we assume that a liquid electrolyte occupies pores within a solid superstructure. We consider an electrolyte comprising LiPF$_6$, EC and a background solvent. Letting $\varepsilon_e$ represent the pore volume fraction, mass continuity is governed by

$$\frac{\partial \varepsilon_e c_+}{\partial t} = -\vec{\nabla} \cdot \vec{N}_+ + R_+^{tot} \tag{1}$$

$$\frac{\partial \varepsilon_e c_{EC}}{\partial t} = -\vec{\nabla} \cdot \vec{N}_{EC} + R_{EC}^{tot} \tag{2}$$

in which $c_{EC}$ is the molar concentration of EC and $c_+$ the molarity of Li$^+$ cations. Local electroneutrality is assumed, so explicit equation for the PF$_6^-$ anion molarity $c_-$ is not needed: under electroneutrality $c_e = c_+ = c_-$, where $c_e$ is the LiPF$_6$ salt concentration. Source terms $R_+^{tot}$ accounts for the contributions of (de-)lithiation and SEI reaction on Li$^+$ cations, whereas $R_{EC}^{tot}$ represents the consumption of EC due to the SEI reaction. Observe that reactions which consume EC are explicitly allowed for in Eq. (1). More detail about the source terms is given below.

A continuity equation governing the concentration $c_0$ of the background solvent is not required. The composition dependence of the total molarity $c_T = c_0 + c_{EC} + c_+ + c_- = c_0 + c_{EC} + 2c_e$ is governed by a thermodynamic equation of state, which under electroneutrality requires that $c_T$ is at most a function of $c_{EC}$ and $c_e$:

$$c_T = \frac{1 - (\bar{V}_{EC} - \bar{V}_0) c_{EC} - (\bar{V}_e - 2\bar{V}_0) c_e}{\bar{V}_0}, \tag{3}$$

where $\bar{V}_{EC}$, $\bar{V}_e$, and $\bar{V}_0$ are the partial molar volume of EC, LiPF$_6$ salt, and background solvent, respectively. If needed, $c_0$ can be calculated if $c_T(c_{EC}, c_e)$ is known, because the particle fractions $y_k = c_k / c_T$ of the ions and cosolvents must sum to unity.

Transport laws that give the fluxes of Li$^+$ and EC in Eqs. (1) and (2) are based on the cosolvent model put forward by Monroe[24]. The total molar fluxes $\vec{N}_i$ of cations and EC are given by

$$\vec{N}_+ = D_e \varepsilon_e \vec{d}_e + \frac{D_\times c_e \varepsilon_e}{c_T} \vec{d}_{EC} + \frac{t_+^0}{F}\vec{\iota}_e + c_e \vec{v}_0 \tag{4}$$

$$\vec{N}_{EC} = \frac{D_\times c_{EC} \varepsilon_e}{c_T} \vec{d}_e + D_{EC} \varepsilon_e \vec{d}_{EC} + \frac{2\Xi}{F}\vec{\iota}_e + c_{EC} \vec{v}_0, \tag{5}$$

where $F$ is Faraday's constant; $\vec{\iota}_e$ is the electrolyte current density. $\vec{v}_0$ is the average velocity of the background solvent; $\vec{d}_e$ and $\vec{d}_{EC}$ are the thermodynamic forces driving diffusion of the neutral-solute components; $D_e$, $D_{EC}$, and $D_\times$ are the thermodynamic electrolyte diffusivity, thermodynamic EC diffusivity, and cross diffusivity, respectively; and $t_+^0$ and $\Xi$ are respectively the cation transference number and EC migration coefficient. Note that the two thermodynamic forces above are expressed with units of molar-concentration for convenience[24]:

$$\vec{d}_e = -\frac{c_T c_e \vec{\nabla} \mu_e}{2c_0 RT} \text{ and } \vec{d}_{EC} = -\frac{c_T c_{EC} \vec{\nabla} \mu_{EC}}{c_0 RT}, \tag{6}$$

where $\mu_e$ and $\mu_{EC}$ are the electrochemical potential of LiPF$_6$ salt and EC, respectively. Under isothermal, isobaric conditions, the thermodynamic driving forces relate to concentration gradients through[24]

$$\vec{d}_e = -\chi_{ee} \vec{\nabla} c_e - \chi_{e,EC} \vec{\nabla} c_e, \tag{7}$$

$$\vec{d}_{EC} = -\chi_{EC,e} \vec{\nabla} c_e - \chi_{EC,EC} \vec{\nabla} c_{EC}. \tag{8}$$

In which the thermodynamic factors $\chi_{ij}$ relate to the activity coefficients $f_i$ through

$$\chi_{ij} = \frac{c_T}{c_0}\left[\delta_{ij} + c_i \left(\frac{\partial \ln f_i}{\partial c_j}\right)_{T,p}\right] \tag{9}$$

where $\delta_{ij}$ is the Kronecker delta. In moderately dilute cases, the activity coefficient $f_i$ varies primarily with $c_i$, so the activity matrix can be assumed diagonal [2], i.e.,

$$\chi_{e,EC} \approx 0, \chi_{EC,e} \approx 0. \tag{10}$$

Neglecting convection ($\vec{v}_0 = 0$) and using a Bruggeman exponent $b$ to account for pore tortuosity, Eqs. (1) and (2) become

$$\vec{N}_+ = -\varepsilon_e^b D_e v_+ \vec{\nabla} c_e - \varepsilon_e^b D_\times \vec{\nabla} c_{EC} + \frac{t_+^0}{F} \vec{i}_e \tag{11}$$

$$\vec{N}_{EC} = -\varepsilon_e^b D_\times \vec{\nabla} c_e - \varepsilon_e^b D_{EC} \vec{\nabla} c_{EC} + \frac{2\Xi}{F} \vec{i}_e \tag{12}$$

**Charge conservation equation in the electrolyte**

Following Ref [24], if we use measured liquid junction potential (LJP), the modified ohmic law with contributions from EC and $Li^+$ concentration gradients is:

$$\vec{i}_e = -\kappa \vec{\nabla} \phi_e + \frac{\partial \Delta U}{\partial c_e} \vec{\nabla} c_e + \frac{\partial \Delta U}{\partial c_{EC}} \vec{\nabla} c_{EC}. \tag{13}$$

For the normal DFN (which we call "single-solvent" case), the above equation reads:

$$\vec{i}_e = -\kappa \vec{\nabla} \phi_e + \frac{\partial \Delta U}{\partial c_e} \vec{\nabla} c_e. \tag{14}$$

For charge conservation, we have:

$$\vec{\nabla} \cdot \vec{i}_e = \frac{a j_+^{tot}}{F} = \begin{cases} \frac{a j_+^{int}}{F} + \frac{a j_+^{SEI}}{F}, & \text{for negative electrode} \\ 0, & \text{for separator} \\ \frac{a j_+^{int}}{F}, & \text{for positive electrode} \end{cases} \tag{15}$$

Substituting Eqs. (13) or (14) into Eq. (15) gives the charge conservation in terms of EC and $Li^+$ concentration.

**SEI reaction with EC concentration dependency**

To further investigate the effect of the double transport model on the degradation of LIBs, the SEI reaction is considered here. We assume the following EC-based two-electron SEI reaction:

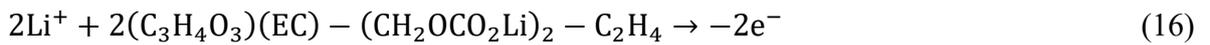

$$2Li^+ + 2(C_3H_4O_3)(EC) - (CH_2OCO_2Li)_2 - C_2H_4 \rightarrow -2e^- \tag{16}$$

which gives the following stoichiometric coefficients:

$$s_+ = 2, \; s_{EC} = 2, \; s_{SEI} = -1, \; n = -2 \tag{17}$$

Ignoring the volume of $C_2H_4$, the volume reduction by $Li^+$ and EC is larger than the volume increase by SEI layers (in this case, $(CH_2OCO_2Li)_2$). That means part of pores in the electrode will become empty due to solvent consumption. Here we assume that such pores will be refilled by an electrolyte reservoir outside the jelly roll, but inside the cell package, instantly. The refilled electrolyte will bring in extra Li$^+$ and EC, which manifests as another source term. If we assume the electrolyte reservoir has the same Li$^+$ and EC concentrations that the jelly roll had at the start, then the source terms for Li$^+$ and EC are:

$$R_+^{\text{refill}} = -aj_+^{\text{SEI}}\left(\frac{s_{\text{EC}}}{s_+}\bar{V}_{\text{EC}} + \bar{V}_+ + \frac{s_{\text{SEI}}}{s_+}\bar{V}_{(\text{CH}_2\text{OCO}_2\text{Li})_2}\right)c_+^0 \tag{18}$$

$$R_{\text{EC}}^{\text{refill}} = -aj_+^{\text{SEI}}\left(\frac{s_{\text{EC}}}{s_+}\bar{V}_{\text{EC}} + \bar{V}_+ + \frac{s_{\text{SEI}}}{s_+}\bar{V}_{(\text{CH}_2\text{OCO}_2\text{Li})_2}\right)c_{\text{EC}}^0 \tag{19}$$

$a$ is the specific surface area. For spherical particles, $a = 3\varepsilon_e/R$, where $R$ is the radius of the electrode particle. $\bar{V}_i$ and $c_i^0$ are the partial molar volume and initial concentration of species $i$, respectively. $j_+^{\text{SEI}}$ is the interfacial SEI current density, with units of A/m$^2$. We assume that the refilling process only occurs in domains that experience the SEI reaction. In our case, this happens only at the negative electrode. We have ignored the fluid dynamics of the refilling process. More details of the solvent consumption and refilling process can be found in Ref [27]. With that said, the source terms for Li$^+$ and EC are:

$$R_+^{\text{tot}} = \begin{cases} \dfrac{aj_+^{\text{int}}}{F} + \dfrac{aj_+^{\text{SEI}}}{F} - \dfrac{aj_+^{\text{SEI}}}{F}\left(\dfrac{s_{\text{EC}}}{s_+}\bar{V}_{\text{EC}} + \bar{V}_+ + \dfrac{s_{\text{SEI}}}{s_+}\bar{V}_{(\text{CH}_2\text{OCO}_2\text{Li})_2}\right)c_{e,0}, \text{ for neg} \\ 0, \text{ for sep} \\ \dfrac{aj_+^{\text{int}}}{F}, \text{ for pos} \end{cases} \tag{20}$$

$$R_{\text{EC}}^{\text{tot}} = \begin{cases} a\dfrac{s_{\text{EC}}}{s_+}\dfrac{j_+^{\text{SEI}}}{F} - \dfrac{aj_+^{\text{SEI}}}{F}\left(\dfrac{s_{\text{EC}}}{s_+}\bar{V}_{\text{EC}} + \bar{V}_+ + \dfrac{s_{\text{SEI}}}{s_+}\bar{V}_{(\text{CH}_2\text{OCO}_2\text{Li})_2}\right)c_{\text{EC},0}, \text{ for neg} \\ 0, \text{ for sep} \\ 0, \text{ for pos} \end{cases} \tag{21}$$

The expression for $j_+^{\text{int}}$ is the same Butler-Volmer expression used in the normal DFN model and can be found in the supplementary material. Now we will look at the detailed expression of $j_+^{\text{SEI}}$. It is well-acknowledged that the SEI reaction is a self-limiting process and normally leads to a capacity fade proportional to the square root of time ($\sqrt{t}$) in the long term. However, it remains a debate whether the exact rate limiting process is solvent diffusion, lithium interstitial diffusion, electron conduction or electron tunnelling. Each of these rate limiting processes can generate one expression of $j_+^{SEI}$. Single et al.[28] and von Kolzenberg et al.[26] have compared different mechanisms and found that the only mechanism that fits their calendar ageing data is the lithium interstitial diffusion limited one:

$$j_+^{SEI} = \frac{c_{int,Li}}{L_{SEI}} \cdot D_{int} F e^{-(\phi_s - \phi_e)} \tag{22}$$

The above equation overwhelms other type of $j_+^{SEI}$ as it capture the effects of potential, current direction (charge or discharge) and SEI thickness simultaneously, which has been observed in multiple works[28, 29]. Based on that, we further assume the SEI current density has a linear dependency on the EC concentration.

$$j_+^{SEI} = \frac{c_{EC}}{\langle c_{EC} \rangle} \cdot \frac{c_{int,Li}}{L_{SEI}} \cdot D_{int} F e^{-(\phi_s - \phi_e)}, \tag{23}$$

where $\langle c_{EC} \rangle$ is the EC concentration at the reference state. Note that we are not intending to present a new mechanism of SEI or fit the ageing model with experiment data. This is just to show how the double-solvent model can change the ageing behaviour of LIBs. There are many studies [5, 7, 9, 10, 30-32] showing the importance of solvent concentration / solvation structure of Li$^+$ on the SEI growth, most of which [5, 7, 9, 30, 31] are in the microscale. Transferring these understandings from microscale to macroscale (cell level) is beyond the scope of this work.

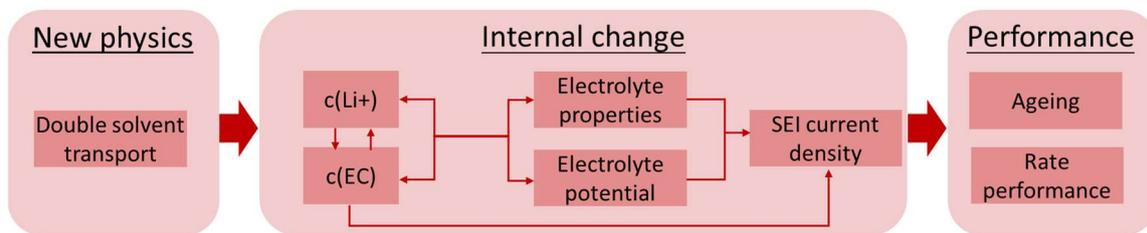

Fig. 1 Summary of the effect of the double-solvent model

Here we summarize how the double-solvent model can affect the cell performance (Fig. 1). Firstly, this model can capture the distribution of EC, which can alter the distribution of Li$^+$ through the cross-diffusion term in Eq. (11). Secondly, quite a few papers [10-13] showed that the electrolyte transport properties are functions of EC concentration. Thirdly, the EC concentration gradient also contributes to the electrolyte potential. Finally, the EC concentration can affect the SEI current density and therefore ageing behaviour.

## Model validation

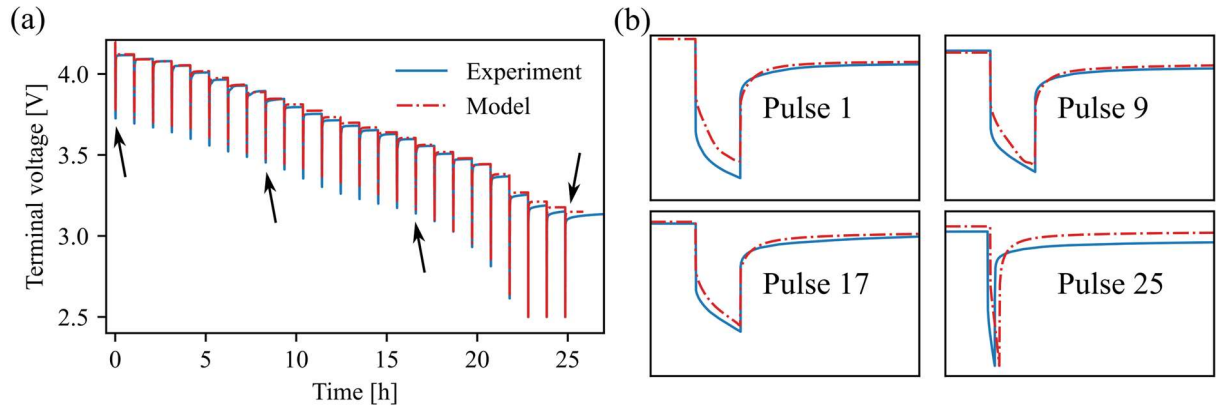

Fig. 2 Model validation using 2C GITT. (a) overall; (b) Zoom of pulses 1,9,17 and 25. (RMSE=116mV)

The model in this work is implemented in the open-source platform PyBaMM[33]. To validate our model, the simulated GITTs were fit to experimental data of the well-parameterized LG M50 cylindrical cell at a current pulse of 1C and 2C. The fully charged cell was subjected to a total number of 25 pulses at 2C, followed by a rest period of 1 hour. Detailed experimental procedures can be found in the Experiment section in Supplementary materials. In Fig. 2(a), the simulated curve fits quite well to the experimental data, especially for the rest periods (flat regions). The overall RMSE is 116mV. In Fig. 2(b), the magnitude of the voltage drops at the beginning of the current pulse mainly reflects the ohmic part of the cell impedance. The simulated voltage drop during the pulse is less than the experimental one for pulse 1, but fits considerably well for pulse 9, 17, and 25. In Fig. S4, the simulated GITT was fit to experimental data at 1C, which also showed good fit with an overall RMSE of 29mV.

However, we need to emphasise that this validation only shows that this model behaves equally well compared with the single-solvent model (normal DFN model, as shown in Fig. S5) in fitting the GITT result. Due to the lack of in situ techniques to quantify the spatial distribution of solvent and Li$^+$ concentration in this well-parametrised commercial cylindrical cell, it is difficult to validate the specific effects of EC concentration gradients directly.

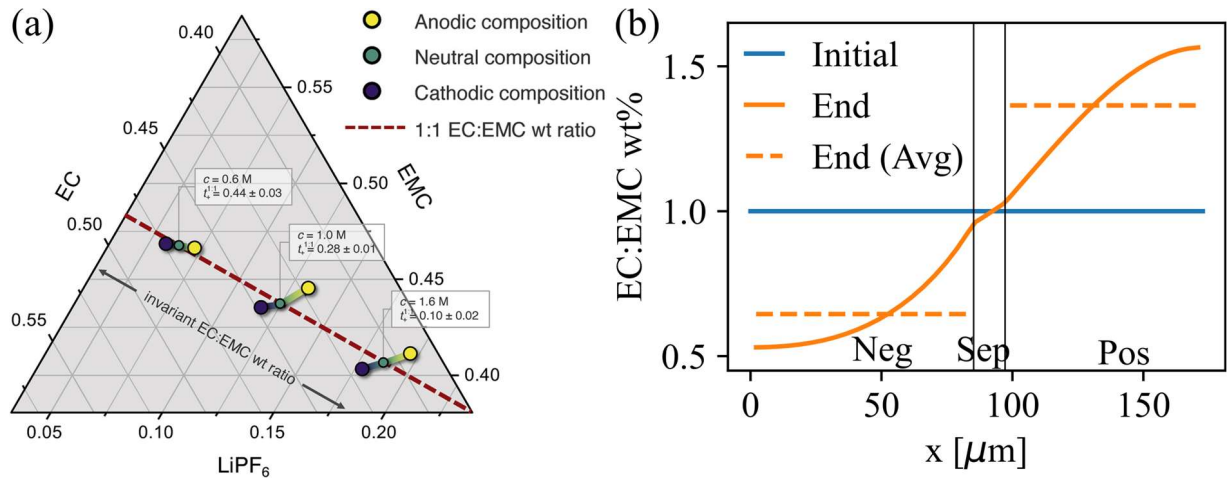

Fig. 3 (a) Experimentally observed solvent segregation[8], (b) modelling results of 3C discharge with 1M initial c(Li$^+$) and EC:EMC=1:1 wt%. (Neg: average: 0.645, minimum: 0.530. Pos: average: 1.365, max: 1.565)

To qualitatively verify our model in a straightforward way, we compare the modelling result in a commercial cell (LG M50) with the experimental observation in a Hittorf cell in Fig. 3. In Fig. 3(a), three electrolytes with same 1:1 EC:EMC mass ratio but salt concentrations of 0.6 M, 1M, 1.6M were polarized in Hittorf cells under an applied current of 0.5 mA (0.64 mAcm$^{-2}$) for 20 hours[8]. The green dots indicate the initial composition whereas the yellow and purple dots indicate the average compositions within the anodic and cathodic chambers of the Hittorf cells at the end of the current pulse, respectively. For the case of 1M salt concentration, the EC:EMC mass ratios were 0.96:1 and 1.03:1 in the anodic and cathodic chambers after the current pulse, respectively. According to the transport model for the Hittorf cell[8], such changes in average concentration result in up to 50% change at the edges of the chamber.

To reproduce this result in the porous electrodes with our double-solvent model, we apply 10A (2C) on a fully charged cell until a voltage cut-off of 2.5V. As presented in Fig. 3(b), the initial EC:EMC mass ratio is 1:1, with a salt concentration of 1M. At the end of 3C discharge, the average EC:EMC mass ratios are 0.645 and 1.365 on the negative and positive electrode, respectively. The maximum solvent segregation occurs at the current collector side, being 0.53 and 1.565 respectively.

The result shows that our double-solvent model can reproduce the experimentally observed solvent segregation. The difference between the average and minimum value is much less than what Wang et al[8]. predicted in their Hittorf cells, mainly due to the large differences in the length of the two domains (0.1728 mm vs 200 mm).

## Model prediction

**Part-1 Internal changes**

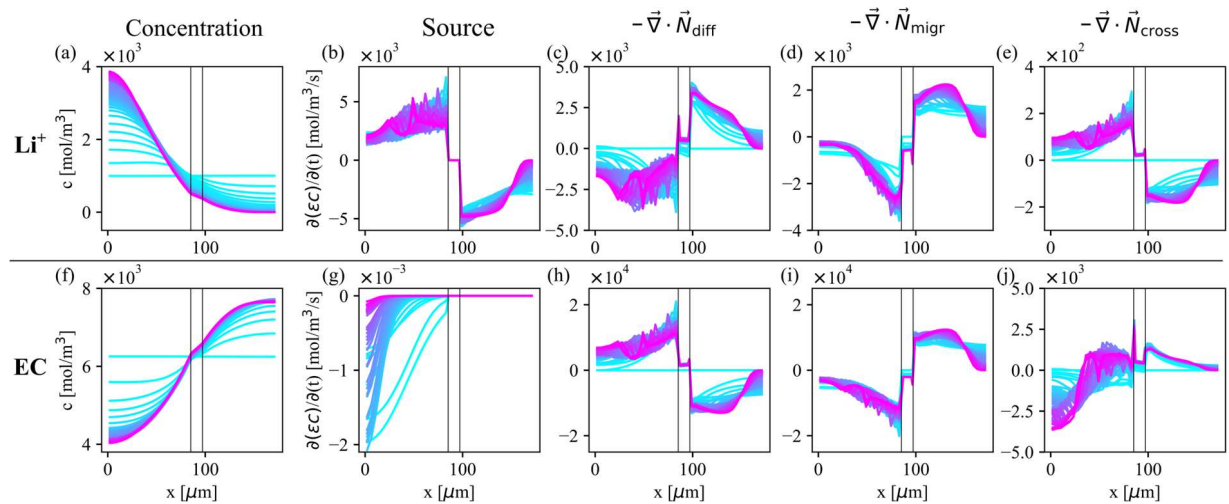

Fig. 4 EC and Li$^+$ concentrations and flux contributions (4.5C discharge, high $D_\times$).

To have a better insight on what this double-solvent model can predict, the EC and Li$^+$ concentration and their contributions during 4.5C discharge are presented in Fig. 4 with high cross diffusivity ($D_\times = 1.5 \cdot 10^{-10} m^2/s$). The results for low $D_\times$ ($D_\times = 1.5 \cdot 10^{-12} m^2/s$) and single-solvent cases are presented in Fig. S8 and Fig. S9, respectively.

The first interesting phenomenon to note in Fig. 4 is that the Li$^+$ and EC have concentration gradients in opposite directions. To be specific, during discharge,, c(Li$^+$) increases at the negative electrode and decreases at the positive electrode (Fig. 4(a)), whereas c(EC) is the other way around (Fig. 4(b)). Both concentration profiles change quickly in the beginning then stabilise towards the end of discharge.

c(Li$^+$) reaches 4M at the end of 4.5C discharge next to the negative current collector, and approaches zero next to the positive current collector, indicating a limiting current due to salt depletion or saturation (at 4M, salt precipitation may also occur [34], which is not included in the model). However, the relative change of c(EC) is less significant, only decreasing by 1/3 next to the negative current collector and increasing by 1/3 next to the positive current collector. This is mainly due to the larger self-diffusivity of EC ($D_{EC}$).

To better understand how the opposite concentration profiles are formed, different driving forces are compared in Fig. 4 (b) ~(e), (g) ~ (j) based on Eqs. (11) and (12). The fluxes are divided into three parts, diffusion, migration, and cross diffusion, each contributing through:

$$-\vec{\nabla} \cdot \vec{N}^+_{\text{diff}} = \varepsilon_e^b D_e v_+ \vec{\nabla} c_e \qquad (24)$$

$$-\vec{\nabla} \cdot \vec{N}^+_{\text{migr}} = -\frac{t_+^0}{F}\vec{i}_e \qquad (25)$$

$$-\vec{\nabla} \cdot \vec{N}^+_{\text{cross}} = \varepsilon_e^b D_\times \vec{\nabla} c_{EC} \qquad (26)$$

$$-\vec{\nabla} \cdot \vec{N}^{EC}_{\text{diff}} = \varepsilon_e^b D_{EC} \vec{\nabla} c_{EC} \qquad (27)$$

$$-\vec{\nabla} \cdot \vec{N}^{EC}_{\text{migr}} = -\frac{2\Xi}{F}\vec{i}_e \qquad (28)$$

$$-\vec{\nabla} \cdot \vec{N}^{EC}_{\text{cross}} = \varepsilon_e^b D_\times \vec{\nabla} c_e \qquad (29)$$

Though different driving forces interplay with each other, it is always convenient to start the analysis from the source term of Li$^+$ as it is the most fundamental driving force. During discharge, Li$^+$ deintercalates from the negative electrode, making a positive source term on the negative side (Fig. 4(b)). Meanwhile, Li$^+$ moves under the electric field from the negative electrode to the positive one, forming a negative "source" in the negative electrode (Fig. 4(d)). However, because the cation transference number is less than 1, the migration flux alone cannot move all the Li$^+$ that deintercalate from the negative electrode to the positive side, so a concentration gradient will build up and contribute a negative "source" in the negative electrode via diffusion (Fig. 4(c)).

For the EC profile, the migrating Li$^+$ ions drag the EC molecules from the negative electrode to the positive electrode (Fig. 4(i)). However, unlike Li$^+$, EC doesn't participate in the interfacial reaction in the positive electrode. Therefore, most EC molecules moving together with Li$^+$ will be left outside the positive electrode. This is why the two species share the opposite distribution. There is a negative EC source term in the negative electrode due to the SEI reaction (Fig. 4(g)). However, this source term is orders of magnitude lower than the others because the SEI reaction is a slow, long-term effect. In other words, the fundamental driving force for EC is migration. It is also worth noting that $\left|-\vec{\nabla}\cdot\vec{N}_{migr}^{EC}\right|$ and $\left|-\vec{\nabla}\cdot\vec{N}_{migr}^{+}\right|$ share a similar shape, but the former is higher because there are normally more EC molecules in one Li$^+$ solvation sheath. Such migration of EC induces an EC concentration gradient and makes a positive "source" via diffusion in the negative electrode (Fig. 4(h)).

Finally, the cross-diffusion term of Li$^+$ depends on the self-diffusion term of EC and therefore has a similar shape, but the same thing doesn't apply the other way around because Li$^+$ has a strong source term due to (de)intercalation. The cross-diffusion term of EC is comparable to the source term of Li$^+$. However, the cross-diffusivity is smaller than the self-diffusivities so the contribution from these cross terms is lower. In Fig. S8 where $D_\times$ is lower, the contribution from the cross terms is even less.

As a comparison, at 1C discharge (Fig. S7), all terms will be much lower except for the source terms for EC due to SEI reaction, as higher SEI current density will occur if c(EC) is higher. However, this source term is still ignorable compared with others.

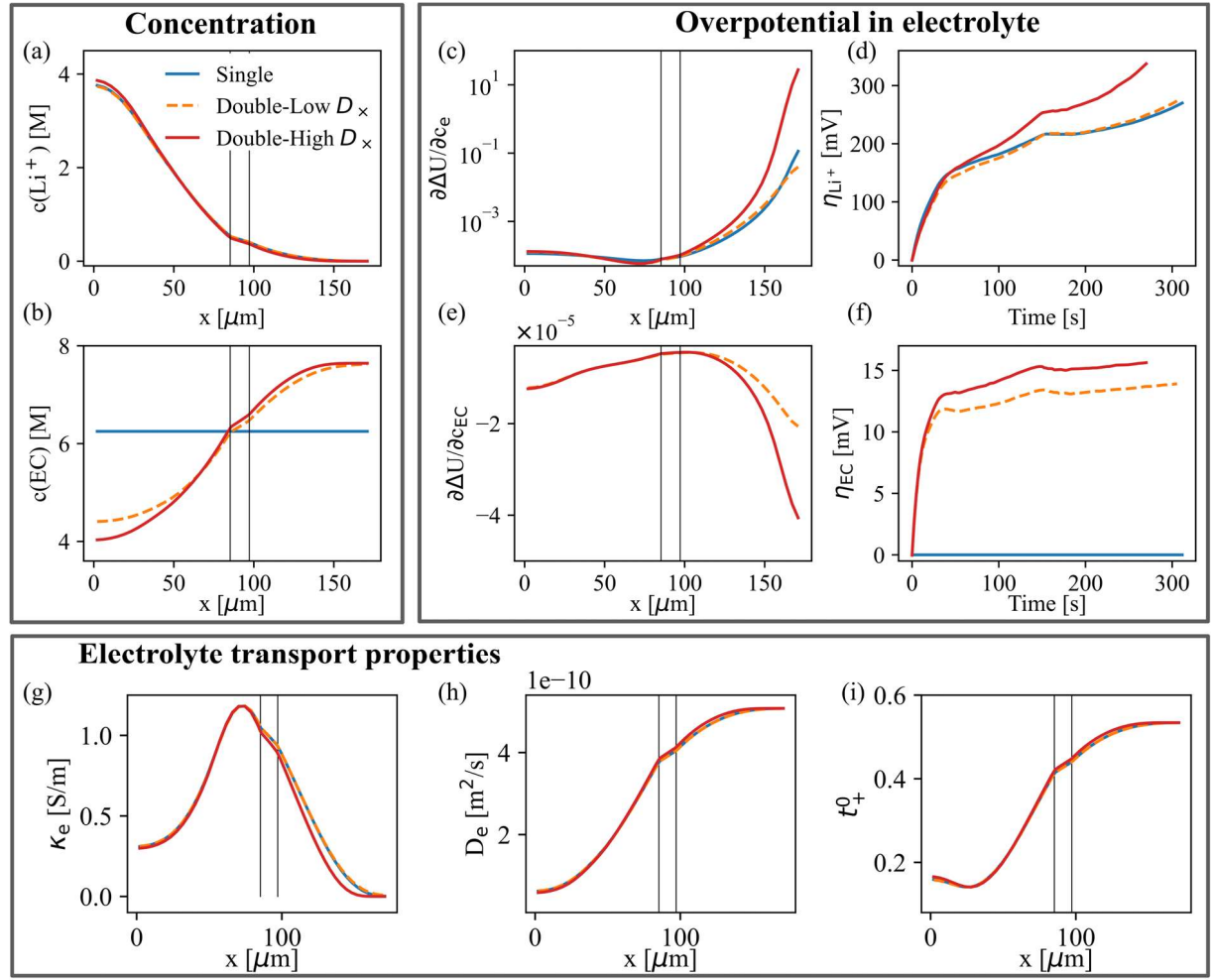

Fig. 5 Internal gradients of concentration, overpotential in electrolyte and electrolyte transport properties at the end of a 4.5C discharge.

The concentration of $Li^+$ and EC distribution in Fig. 4 will further affect the electrolyte properties, electrolyte overpotential, and many other internal states of the model cells. In Fig. 5, we compare the two concentration profiles, the two concentration-induced overpotentials in the electrolyte and the three transport properties as a function of location at the end of a 4.5C discharge. The other three transport properties, i.e., EC transference number, EC diffusivity, and cross diffusivity remain constant, as seen in Table S2. We consider three cases here, the single-solvent case (normal DFN model), and two double solvent cases with different values of $D_\times$.

The c(Li$^+$) profile of the high $D_\times$ case in Fig. 5(a) is approaching 4M and slightly higher than the other two cases at the negative current collector, as the cross-diffusion term increases the c(Li$^+$) concentration gradient (Fig. 4). Accordingly, c(Li$^+$) for the high $D_\times$ case should be lower than the other two cases at the positive current collector, though the differences are indistinguishable as they are all approaching zero. However, the low $D_\times$ case shows the same c(Li$^+$) profile as the single case, confirming that the cross diffusion is the direct way that the double-solvent model can change c(Li$^+$).

The differences in the c(EC) profiles are remarkable. c(EC) is constant for the single case by definition, whereas the double cases show considerable gradient. The concentration gradient in the high $D_\times$ case is higher than the low $D_\times$ one due to the cross diffusion.

The concentration gradient of Li$^+$ and EC will cause overpotential in the electrolyte (Fig. 5(c)~(f)). As presented in Fig. 5(c) and (e), $\frac{\partial \Delta U}{\partial c_e}$ and $\frac{\partial \Delta U}{\partial c_{EC}}$ are more sensitive to the concentration when c(Li+) approaches zero and c(EC) approaches 8 M. Note that logarithmic scale is used for $\frac{\partial \Delta U}{\partial c_e}$ in Fig. 5(c) is it increases by two orders of magnitude next to the positive current collector. The corresponding overpotential can be up to 340 mV for Li$^+$ and 15 mV for EC for the high $D_\times$ case.

The three transport properties for the three cases coincide in Fig. 5(g)-(i) except for the electrolyte conductivity next to the positive current collector.

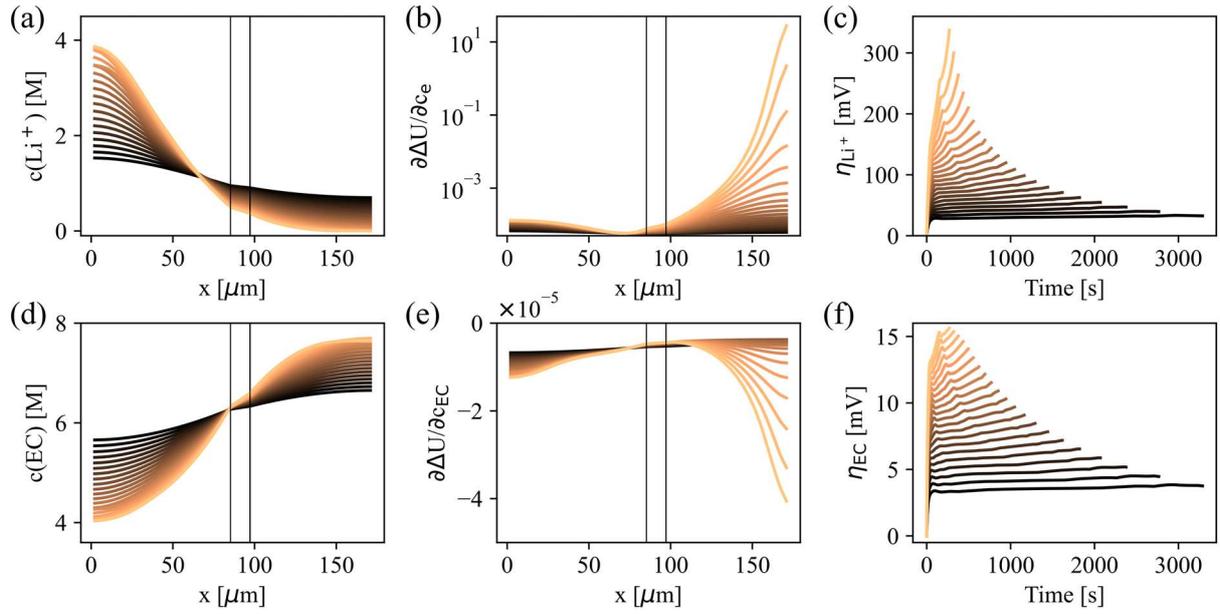

Fig. 6 Concentration profiles and overpotentials in electrolyte for the high $D_\times$ case at different discharge rates.

(C rates start from 1 and increase by 0.175 until 4.5C.)

We further investigate the effect of C-rate on the internal states during discharge. The internal states for the high $D_\times$ case are shown in Fig. 6. (The low $D_\times$ case and single case are presented in Fig. S10 and Fig. S11, respectively.) The applied current rate is between 1C and 4.5C with 0.175 intervals. Both Li$^+$ and EC concentration gradients increase linearly up to 3.1C. After that, both concentration gradients eventually reach their limit, which indicates a limiting current. Accordingly, the absolute values of both $\frac{\partial \Delta U}{\partial c_e}$ and $\frac{\partial \Delta U}{\partial c_{EC}}$ increase with C rate slowly in the beginning but surge after 3.8C. The logarithmic scale is applied to $\frac{\partial \Delta U}{\partial c_e}$ again in Fig. 6(b) to adapt to the quick changing rate. These in turn lead to the non-linear change in both Li$^+$ and EC overpotentials.

**Part-2 Rate performance and cycle life**

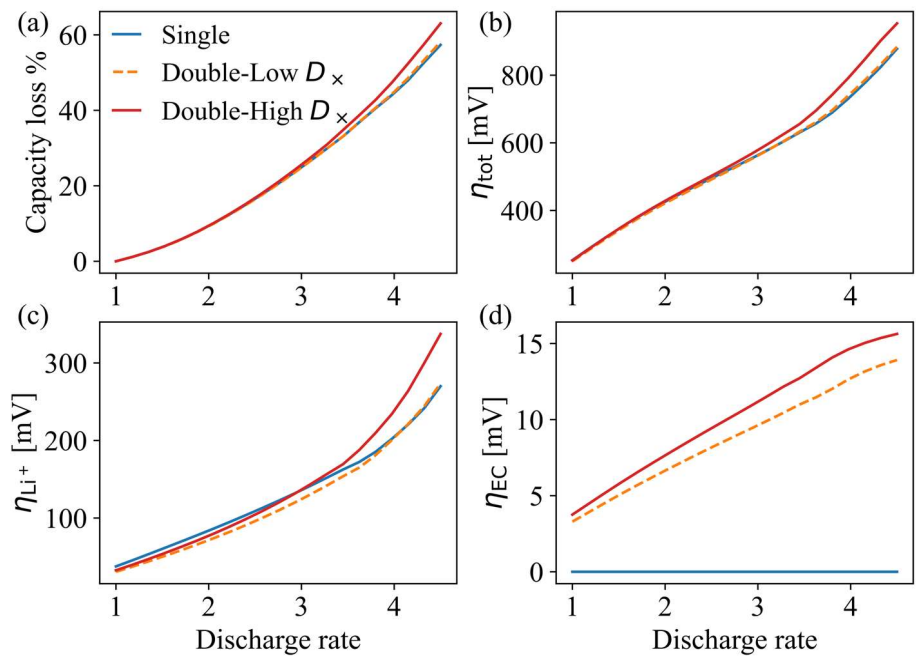

Fig. 7 (a) Rate performance, (b) total overpotential, (c) Li+ overpotential in electrolyte, (d) EC overpotential in electrolyte at the end of discharge

All the internal states mentioned above will ultimately change the performance of LIBs. In Fig. 7, the effect of C-rate on the discharge capacity for the three cases are compared, together with the total overpotential from all sources and the Li$^+$ and EC overpotentials in the electrolyte. The high $D_\times$ case does not show observable difference to the single case until 3.275C. After that, the gap between them quickly increases. At 4.5C, the high $D_\times$ case predicts 6% more capacity loss and 75 mV higher total electrolyte overpotential than the single case. The low $D_\times$ case, however, coincide with the single case for both capacity and total overpotential in the electrolyte when the C rate is less than 4C. This is because the lower Li$^+$ overpotential compensates the higher EC overpotential for the low $D_\times$ case. The differences observed in total overpotential (Fig. 7(b)) are ascribed to the Li$^+$ overpotential, as it takes up to ~40% of the total overpotential at 4.5C.

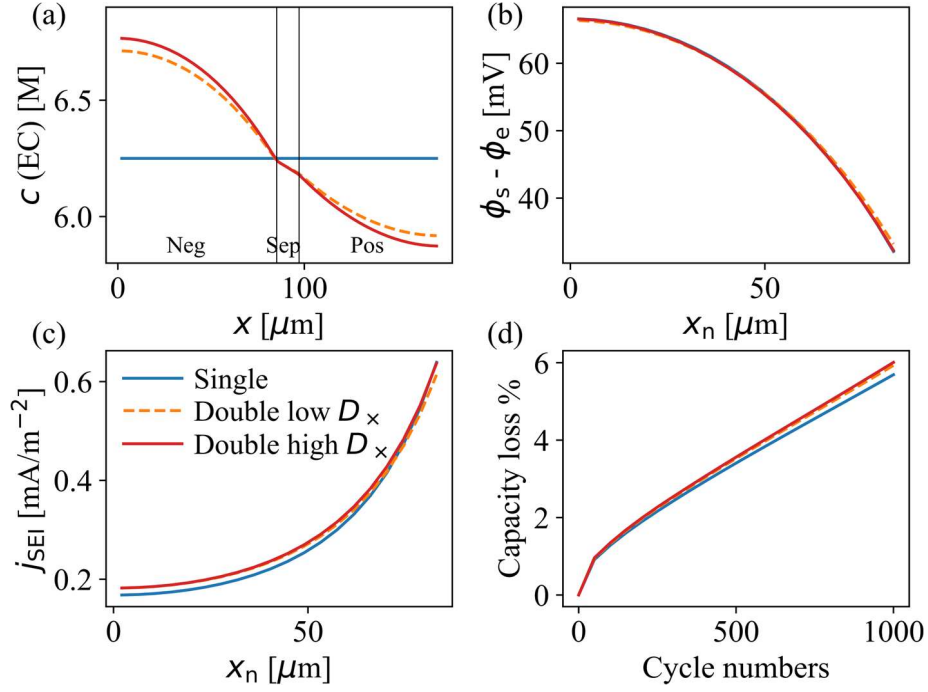

Fig. 8 (a) EC concentration across the cell, (b) solid and liquid potential difference and (c) SEI interfacial current density at the negative electrode at the end of 1C charge for the beginning of life (BOL). (d) The resulting capacity loss after 1000 cycles of 1C. All three cases consider solvent consumption and electrolyte replenishment.

We now investigate how the double-solvent model affects the ageing behaviour of LIBs. We compare the SEI interfacial current density at BOL and capacity loss after 1000 cycles at 1C in Fig. 8. All three cases consider solvent consumption and electrolyte replenishment. Based on Eq. (23), the SEI current density depends on the local EC concentration (Fig. 8(a)) and solid and liquid potential difference ($\phi_s - \phi_e$) (Fig. 8(b)). Both the double cases show considerable c(EC) gradient, with the maximum and minimum values occurring at the current collectors. The c(EC) differences are much less next to the separator, with all three cases coinciding at the negative-separator interface. $\phi_s - \phi_e$ is almost the same for the three cases throughout the negative electrode. Moreover, $\phi_s - \phi_e$ has a bigger impact on the SEI current density ($j_{SEI}$) than c(EC) as it manifests an exponential term while c(EC) only holds a linear term. Therefore, even though the double-high $D_\times$ case shows slightly higher $j_{SEI}$ than the single case at the negative current collector, the difference eventually diminishes and disappears at the negative-separator interface in Fig. 8(c). The overall effect will be lower considering the highest $j_{SEI}$ occurs at the negative-separator interface. The double-low $D_x$ case even shows slightly less $j_{SEI}$ than the single case at the negative-separator interface due to higher $\phi_s - \phi_e$. As a result, the two double cases only show ~0.32% more capacity loss after 1000 1C cycles.

From Li et al.[27], we know that if solvent consumption is considered, the solvent concentration will eventually decrease during ageing, leading to lower SEI growth rate. If this mechanism is ignored, all three cases would predict more capacity loss (Fig. S16). The differences between the single and double cases would become ~0.62%.

## Discussion on the electrolyte properties

### Concentration ranges during measurement

There are two difficulties in battery modelling when it comes to electrolyte properties. The first one, as mentioned in the Introduction section, is the large number of electrolyte properties required if we discard the single-solvent assumption. The second one is the concentration range where these properties are measured. It is expensive and difficult to measure some properties at extreme concentration, especially when approaching the salt saturation limit, as the complex precipitation/dissolution dynamics step in. Moreover, the number of samples required will be large if $Li^+$, EC, and temperature all change simultaneously.

To be used in the battery models, the processed data will normally be interpolated with analytic functions. This works well if the model operates within the measurement range. However, many state-of-the-art studies tend to push the electrolyte properties outside their measurement ranges, especially during rate performance or long-term ageing simulations, even though many modelling papers do not show the electrolyte properties explicitly. Therefore, it becomes critical to select the right form of the fitting functions if extrapolation is applied, so that the modelling results are still reasonable. Moreover, to keep the result genuine, special notices should be given when the model operates outside the measurement ranges of all parameters, not just the electrolyte properties.

Nyman et al.[35] and Valøen et al.[36] measure the electrolyte conductivity and diffusivity up to 2.2 M and 3.3 M, respectively, and fit the data with polynomial functions. Therefore, as shown in Fig. S15, unrealistically high conductivity will be predicted if $c(Li^+)$ is higher than the measurement range. Landesfeind and Gasteiger[11] chose the exponential form for the electrolyte properties and measure them at $c(Li^+)$ of up to 4M, which ensures low conductivity and diffusivity at higher $c(Li^+)$. Therefore, the electrolyte diffusivity, conductivity and cation transference number in this work are chosen from Landesfeind and Gasteiger[11].

In Jung et al.[23], the measurement range for EC and Li are $0 < y_{EC} < 0.75$ and $0.002 < y_e < 0.15$, respectively. For our rate performance study, $y_e$ exceeds 0.15 at the negative current collector above 2.05C, whereas $y_e$ becomes lower than 0.002 at the positive current collector above 3.975C. $y_{EC}$ never exceeds the measurement range in all the cases in this work. Below 2.05C, the double-solvent model does not show observable differences with the single case except for the ageing results (Fig. 8). If we only avoid results exceeding the lower bound of $y_e$, the highest C-rate to study is 3.8C, where the double-high $D_\times$ case predicted 2.3% more power fade and 53 mV total overpotential in the electrolyte. Therefore, the effect of the double-solvent model is still significant.

**Three diffusivities in the electrolyte**

The magnitude of the three diffusivities ($Li^+$, EC, and cross diffusivities) dominates the concentration gradient of both $Li^+$ and EC. The lower the diffusivity, the higher the concentration gradient. Therefore, to ensure genuine results, the magnitude of the three diffusivities is discussed here. $Li^+$ diffusivity has been more extensively measured compared to the other two diffusivities. In this work, we use the well-characterized $Li^+$ diffusivity from

Landesfeind and Gasteiger[11]. Cross diffusivity, to our best knowledge, has not been measured properly before. However, a range can be given based on thermodynamic stability as[24]:

$$0 \leq D_\times < \frac{D_e \langle c_T \rangle}{2 \langle c_e \rangle} \tag{30}$$

Based on Fig. S3(a), we approximate $D_e$ with $3 \cdot 10^{-10} m^2/s$. Combining this with $\langle c_T \rangle = 13484.224\ mol/m^3$, we get $0 \leq D_\times < 2.02 \cdot 10^{-9} m^2/s$. However, we found that the model can easily run into errors if the cross diffusivity is larger than the two self-diffusivities, as the cross-diffusion term will enlarge the concentration gradients of either species (Fig. 4). Therefore, we set $D_\times$ to be $1.5 \cdot 10^{-12} m^2/s$ and $1.5 \cdot 10^{-10} m^2/s$ to investigate the cases with weak and strong cross diffusion, respectively.

As for EC diffusivity, previous experimental[37-39] and modelling[40] works have proven that it should be higher than Li$^+$ diffusivity. This is because Li$^+$ move in a solvation cluster that contains not just Li$^+$ but also several solvent molecules (preferentially EC). The dynamic size of such Li$^+$ clusters is larger than those of uncoordinated EC molecules. In this work, most of the results are obtained with $D_{EC} = 5 \cdot 10^{-10} m^2/s$, which comes from Uchida and Kiyobayashi[6]. The sensitivity of $D_{EC}$ is studied by sweeping three values of $D_{EC}$ in double-solvent cases with low $D_\times$ at 4.5C discharge (Fig. S12). As presented in Fig. S12(a)~(c), the concentration and resulting overpotential for Li$^+$ in the electrolyte is almost unaffected by $D_{EC}$. In Fig. S12(d), the EC concentration gradient increases as $D_{EC}$ decreases. The resulting EC overpotential also increases from 9 mV to 21 mV. Note that for $D_{EC} = 3 \cdot 10^{-10} m^2/s$, it is larger than $D_e$ for $c_e > 800\ mol/m^3$ and therefore should only be regarded as the low boundary of $D_{EC}$ in this work. However, even with the lowest possible $D_{EC}$, c(EC) is far from depleted (as seen in the negative electrode in Fig. S12(d)), indicating that solvent being depleted won't be a reason for the limiting current in LIBs as in fuel cells [41].

**Electrolyte conductivity**

We are aware that Wang et al.[8] also measured the conductivity of EC:EMC in LiPF$_6$ with varying compositions. However, a proper fit with an explicit expression for $\kappa_e$ in terms of of $c_e, c_{EC}, T$ is lacking. Moreover, due to the limited concentration range (maximum $c_e$ being slightly higher than 2M), the ability of the model to explore high C-rates or ageing cases is be limited. Nonetheless, we use a fourth-order polynomial function to fit the measurement in Wang et al.[8] as follows:

$$x = \begin{cases} (c_e - 971.4)/733.1, & c_e < 2300 \\ (2300 - 971.4)/733.1, & c_e \geq 2300 \end{cases} \tag{31}$$

$$y = (c_{EC} - 4737)/3439 \tag{32}$$

$$\kappa_e = 0.9608 + 0.1502 \cdot x + 0.173y - 0.3934x^2 - 0.1179xy - \tag{33}$$
$$0.1472y^2 + 0.04244x^3 - 0.1197x^2y - 0.003226xy^2 +$$
$$0.01664x^4 + 0.05145x^3y + 0.05983x^2y^2$$

Eq. (33) fits the measured data with $R^2 = 0.96$. However, the fitting function starts to increase for $c_e \geq 2300$ mol/m$^3$. To avoid possible artifacts, we have to assume constant extrapolation beyond that range (Fig. S13).

As presented in Fig. S13, within the measured range (0~2300 mol/m$^3$), the electrolyte conductivity is similar at different EC concentrations, except for the c(EC)=3M case. Therefore, we found that all the modelling results presented in the main text will be almost unchanged by using the conductivity in Eq. (33), as indicated in Fig. S14.

## Conclusion

Extensive experimental and modelling works have shown that the single-solvent assumption of the DFN model is invalid and that the distribution of solvent concentration has a significant impact on the performance of LIBs. In order to retrieve this behaviour, it is necessary to upgrade the traditional DFN model to include two solvents. A double-solvent model (two solvent, one salt) is proposed and implemented in this work. EC is selected as the "working" solvent due to its preference in the Li$^+$ solvation sheath.

The double-solvent model is validated quantitatively with GITT in a well-parameterised commercial cell (LG M50) and qualitatively with the experimentally observed solvent segregation results in a Hittorf cell. With the validated model, the distribution of EC and Li$^+$ and their contributions can be predicted. The two species show opposite concentration profiles, mainly because Li$^+$ can (de-)intercalate (from) into the active material but EC cannot.

The double-solvent model affects the performance of LIBs by changing the species concentrations, transport properties, and concentration overpotentials in the electrolyte. The high $D_\times$ case has the highest concentration gradient and therefore the highest Li$^+$ and EC concentration overpotentials. The corresponding conductivity approaches zero at the positive electrode at the end of a 4.5C discharge for the virtual LGM50 cell due to zero Li$^+$ concentration. As a result, the high $D_\times$ case predicts 6% less useable capacity at 4.5C (compared to 1C) than the single case.

The double-solvent model is further coupled with an interstitial-diffusion limited SEI growth model with a linear dependency on EC concentration. The model shows limited effects on the ageing performance, predicting only ~0.32% more capacity loss compared to the single-solvent case. This is because the largest difference in EC concentration occurs at the negative current collector, whereas most SEI growth occurs near the separator where the potential difference between the solid and liquid phase ($\phi_s - \phi_e$) is smaller. We assume the SEI current density to have an exponential dependence[26] on $-(\phi_s - \phi_e)$, which is almost the same for the three cases throughout the negative electrode. More complicated interactions between EC transport and SEI growth could be introduced based on findings from the microscale.

In fuel cells, the current can be limited by the dry-out of membranes due to electro-osmotic drag. However, the equivalent analogy of solvent dry-out doesn't apply in LIBs with organic electrolytes. This is because even with the lowest possible EC diffusivity ($3 \cdot 10^{-10} m^2/s$), c(EC) is far from depleted during cycling. Even if the concentrations of both EC and the other solvent reduce significantly, salt precipitation will occur before completely dry-out.

Note that the only parameter used in the model as a function of both EC and Li$^+$ concentration is the measured liquid junction potential used in the MacInnes equation. In the future, it will be interesting to see how the model performs with more key electrolyte parameters being functions of both solvent and salt concentration. Moreover, the electrode parameters in this work correspond to an energy-type LIB (LG M50), with the recommended discharge rate being only 1C. The double-solvent model may deviate more from the single-solvent model for other cells, perhaps significantly for power-type LIBs.

## Conflicts of interest

The authors have no conflicts of interest to report.

## Author statement

**Ruihe Li**: Conceptualization, Software, Formal analysis, Visualization, Writing – Original Draft. **Simon O'Kane**: Validation, Writing – Review & Editing, Supervision. **Andrew Wang**: Resources, Data Curation. **Taeho Jung**: Resources, Data Curation. **Niall Kirkaldy**: Resources, Data Curation. **Monica Marinescu**: Investigation, Supervision. **Charles Monroe**: Methodology, Supervision. **Gregory J Offer**: Conceptualization, Writing – Review & Editing, Supervision, Funding acquisition.

## Acknowledgement

The authors would like to acknowledge financial support from EPSRC Faraday Institution Multiscale Modelling project (EP/ S003053/1, grant number FIRG059). The first author is funded as a PhD student by the China Scholarship Council (CSC) Imperial Scholarship. We also thank Mr. Jianbo Huang (GitHub: huang-b) for the kind help on debugging the Python scripts for this work, and Drs. Valentin Sulzer and Robert Timms for the kind help on implementing the double-solvent model in PyBaMM.

## Supplementary materials

### Other equations in the model

Table S1. All other equations follow the original work by Doyle et al. and are presented here for the authors' convenience.

| Description | Equation | Boundary condition |
|---|---|---|
| | Electrode | |
| Mass conservation | $\dfrac{\partial c_{s,k}}{\partial t} = \dfrac{1}{r^2}\dfrac{\partial}{\partial r}\left(r^2 D_{s,k}\dfrac{\partial c_{s,k}}{\partial r}\right)$ | $\left.\dfrac{\partial c_{s,k}}{\partial r}\right|_{r=0} = 0,\ -D_{s,k}\left.\dfrac{\partial c_{s,k}}{\partial r}\right|_{r=R_k} = \dfrac{j_k^{\text{tot}}}{a_k F}$ |
| Charge conservation | $\dfrac{\partial}{\partial x}\left(\sigma_{s,k}\dfrac{\partial \phi_{s,k}}{\partial x}\right) = j_k^{\text{tot}}$ | $-\sigma_{s,n}\left.\dfrac{\partial \phi_{s,n}}{\partial x}\right|_{x=0} = -\sigma_{s,p}\left.\dfrac{\partial \phi_{s,p}}{\partial x}\right|_{x=L} = i_{app}$  $-\sigma_{s,n}\left.\dfrac{\partial \phi_{s,n}}{\partial x}\right|_{x=L_n} = -\sigma_{s,p}\left.\dfrac{\partial \phi_{s,p}}{\partial x}\right|_{x=L-L_p} = 0$ |
| | Reaction kinetics | |
| Butler-Volmer | $j_k^{\text{int}} = \begin{cases} a_k j_{0,k}^{\text{int}} \sinh\left(\dfrac{1}{2}\dfrac{RT}{F}\eta_k^{\text{int}}\right), & k \in \{n,p\}, \\ 0, & k = s. \end{cases}$ | |
| Overpotential | $\eta_k^{\text{int}} = \begin{cases} \phi_{s,n} - \phi_{e,n} - U_n^{\text{OCV}}\left(c_{s,n}|_{r=R_n}\right) - \eta_{\text{SEI}}, & k = n, \\ \phi_{s,p} - \phi_{e,p} - U_p^{\text{OCV}}\left(c_{s,p}|_{r=R_p}\right), & k = p. \end{cases}$ | |
| | Initial conditions | |
| Initial conditions | $c_{s,k} = c_{k0}\ (k=n,p),\ c_{e,k} = c_{e0}\ (k=n,s,p)$ | |
| | Terminal voltage | |
| Terminal voltage | $V = \phi_{s,p}\big|_{x=L} - \phi_{s,n}\big|_{x=0}$ | |

**Experiment**

GITT experiments were performed on LG M50T cells (LG INR21700-M50T). Cells were first fully charged by a constant current-constant voltage method at 0.3C to 4.2 V, with a C/100 cut-off, followed by a 2-hour rest. 25-pulse GITT tests were performed with discharge pulses of 1C or 2C, which each passed 200 mAh of charge (pulse duration of 144 seconds at 1C or 77 seconds at 2C), with 1 hour rest periods between pulses. Each pulse also had a lower voltage termination condition of 2.5 V to prevent overdischarge. For C-rate determination, the nominal capacity of 5 Ah was used (i.e., 1C equalled 5 A). All tests were performed at a temperature of 25°C. Full details of the experimental procedures, apparatus, and thermal management can be found in Kirkaldy et al. [42]

**Model parameters**

The parameters used in this work are based on a well parameterised cell LG M50. This cell was first parameterised by Chen et al.[43] then improved by O'Regan et al.[44] Apart from these, the EC transport related parameters are mainly from Wang et al.[8] The measured liquid junction potentials are from Jung et al.[23] Though this work does not consider the temperature effect, the Arrhenius dependency of some parameters are included for the readers' convenience. The reference temperature in the Arrhenius relation is always 298.15 K.

Table S2. The parameters used for the DFN model in this study.

| Type | Parameter | Unit | Positive electrode | Separator | Negative electrode |
|---|---|---|---|---|---|
| | Active material | | $Li_xNi_{0.8}Mn_{0.1}Co_{0.1}O_2$ | Ceramic coated polyolefin | $Li_xC_6 + SiO_2$ |
| Design specifications | Current collector thickness | m | $1.6 \cdot 10^{-5}$ | | $1.2 \cdot 10^{-5}$ |
| | Electrode thickness ($L$) | m | $7.56 \cdot 10^{-5}$ | $1.2 \cdot 10^{-5}$ | $8.52 \cdot 10^{-5}$ |
| | Electrode length | m | 1.58 | | |
| | Electrode width | m | $6.5 \cdot 10^{-2}$ | | |
| | Mean particle radius ($R$) | m | $5.22 \cdot 10^{-6}$ | | $5.86 \cdot 10^{-6}$ |
| | Electrolyte volume fraction ($\varepsilon_e$) | | 0.335 | 0.47 | $0.240507^e$ |
| | Active material volume fraction ($\varepsilon_s$) | | 0.665 | | 0.75 |
| | Bruggeman exponent ($b$) | | 1.5 | 1.5 | 1.5 |
| Electrode | Solid phase lithium diffusivity ($D_s$) | $m^2 \cdot s^{-1}$ | Eqs. (S5), Table S3 | | Eq. (S6), Table S3 |
| | Solid phase electronic conductivity ($\sigma_s$) | $S \cdot m^{-1}$ | 0.8473 | | 215 |
| | Maximum concentration ($c_s^{max}$) | $mol \cdot m^{-3}$ | $52787^e$ | | $32544^e$ |
| | Stoichiometry at 100% SOC | | $0.23553^e$ | | $0.88413^e$ |
| Electrolyte | $Li^+$ diffusivity in the electrolyte ($D_e$) | $m^2 \cdot s^{-1}$ | | EC:DMC 1:1 wt% in $LiPF_6$, Eq. (S8) | |
| | Electrolyte ionic conductivity ($\kappa_e$) | $S \cdot m^{-1}$ | | EC:DMC 1:1 wt% in $LiPF_6$, Eq. (S9) | |
| | EC diffusivity in the electrolyte ($D_{EC}$) | $m^2 \cdot s^{-1}$ | | $H(c_{EC}) \cdot 5 \cdot 10^{-10}$ | |
| | Cross diffusivity ($D_\times$) | $m^2 \cdot s^{-1}$ | | $H(c_{EC}) \cdot 1.5 \cdot 10^{-10}$ or $H(c_{EC}) \cdot 1.5 \cdot 10^{-12}$ | |
| | Cation transference number ($t_+^0$) | | | EC:DMC 1:1 wt% in $LiPF_6$, Eq. (S10) | |

| | | | | |
|---|---|---|---|---|
| | Measured liquid junction potential ($\Delta U$) | V | | EC:EMC (varying wt% ratio) in LiPF$_6$, Eq. (S11) for the double-solvent model and Eq. (S17) for the single-solvent case |
| | Migration coefficient ($\Xi$) | | | $0.85 \dfrac{c_{EC}}{\langle c_{EC} \rangle}$ |
| | Initial Li$^+$ concentration in electrolyte ($c_{e,0}$) | mol·m$^{-3}$ | | 1000 |
| | Initial EC concentration in electrolyte ($c_{EC,0}$) | mol·m$^{-3}$ | | 6250 |
| | Reference Li$^+$ concentration in electrolyte ($\langle c_e \rangle$) | mol·m$^{-3}$ | | 1000 |
| | Reference EC concentration in electrolyte ($\langle c_{EC} \rangle$) | mol·m$^{-3}$ | | 6250 |
| Intercalation reaction | Open Circuit Voltages ($U_i^{OCV}$) | V | Eq. (S1) | Eq. (S2) |
| | Exchange current density ($j_{0,k}^{int}$) | A·m$^{-2}$ | Eq. (S3) | Eq. (S4) |

e: estimated from parameter fitting.

Specifically, the initial capacity in the negative electrode is tuned to fit the initial capacity of the GITT. The actual capacity of the cell we tested for GITT is 4.85 Ah at C/10, indicating minor capacity loss. The initial SEI thickness is changed accordingly, see Table S4. $H(c_{EC})$ indicates the Heaviside step function, which is to avoid EC concentration going negative. Note that the electrolyte properties above are from multiple references because no single paper in the current literature has measured a complete set of electrolyte transport and thermodynamic properties assuming double solvents. That would require all properties being functions of both salt and solvent concentrations, which would be a massive experimental effort. Therefore, the only property changing with both c(Li$^+$) and c(EC) is the measured liquid junction potential and the resulting derivatives.

Electrode OCVs are from unpublished measurements taken by Dr. Kieran O'Regan using the methods of Chen et al.[38].

$$U_p^{OCV} = -0.7983 \cdot x + 4.513 - 0.03269 \cdot \tanh(19.83 \cdot (x - 0.5424)) - 18.23 \quad (S1)$$
$$\cdot \tanh(14.33 \cdot (x - 0.2771)) + 18.05 \cdot \tanh(14.46 \cdot (x - 0.2776))$$

$$U_n^{OCV} = 1.051 \cdot e^{-26.76 \cdot x} + 0.1916 - 0.05598 \cdot \tanh(35.62 \cdot (x - 0.1356)) - 0.04483 \quad (S2)$$
$$\cdot \tanh(14.64 \cdot (x - 0.2861)) - 0.02097 \cdot \tanh(26.28 \cdot (x - 0.6183))$$
$$- 0.02398 \cdot \tanh(38.1 \cdot (x - 1))$$

Exchange current density for intercalation is:

$$j_{0,p}^{int} = 3.42 \cdot 10^{-6} \cdot e^{-\frac{1.78 \cdot 10^4}{R}\left(\frac{1}{T}-\frac{1}{298.15}\right)} \cdot c_e^{0.5} \cdot c_{s,suf}^{0.5} \cdot (c_{s,max} - c_{s,suf})^{0.5} \quad (S3)$$

$$j_{0,n}^{int} = 2.668 \cdot e^{-\frac{4 \cdot 10^4}{R}\left(\frac{1}{T}-\frac{1}{298.15}\right)} \cdot \left(\frac{c_e}{\langle c_e \rangle}\right)^{0.208} \cdot \left(\frac{c_{s,suf}}{c_{s,max}}\right)^{0.792} \cdot \left(1 - \frac{c_{s,suf}}{c_{s,max}}\right)^{0.208} \quad (S4)$$

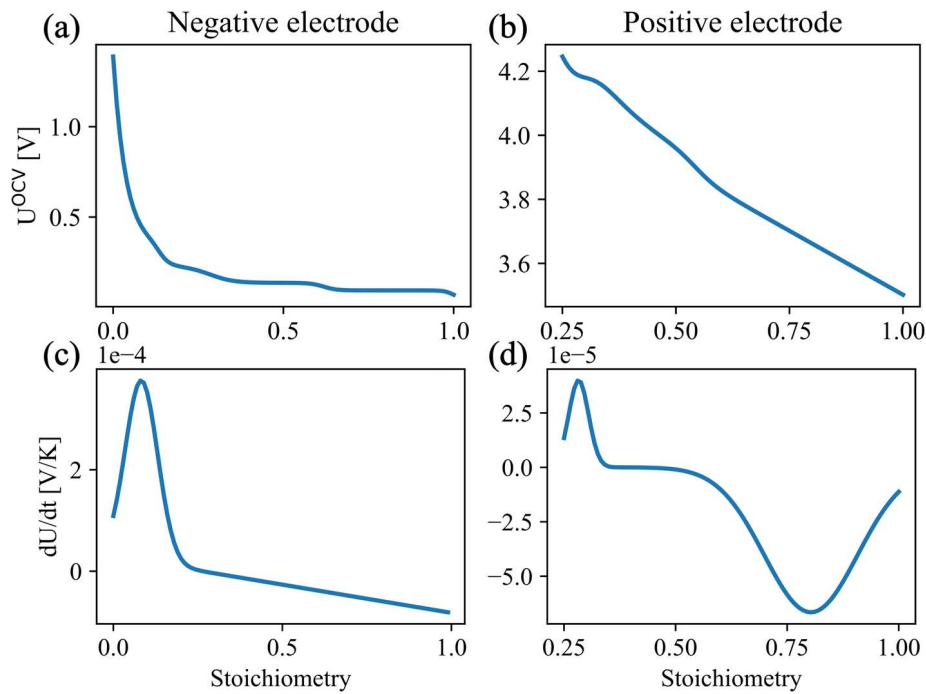

Fig. S1 (a)~(b) OCV and (c)~(d) entropy change[44] of both electrodes.

Solid-phase diffusivity for negative and positive electrode follows:

$$\log_{10}(D_s^{\text{ref}}/R_{\text{cor}}) = a_0 \cdot x + b_0 + a_1 \cdot e^{-\frac{(x-b_1)^2}{c_1}} + a_2 \cdot e^{-\frac{(x-b_2)^2}{c_2}} + +a_3 \cdot e^{-\frac{(x-b_3)^2}{c_3}} + a_4 \cdot e^{-\frac{(x-b_4)^2}{c_4}} \tag{S5}$$

$$D_s = D_s^{\text{ref}} \cdot e^{-\frac{E_{\text{act}}}{R} \cdot \left(\frac{1}{T} - \frac{1}{T_{\text{ref}}}\right)} \tag{S6}$$

Table S3. Fitting parameters for the function describing solid-phase electrode diffusivity[44]. A "-" means that the term including that parameter has not been included.

| Fitting Parameter | Positive Electrode | Negative Electrode |
|---|---|---|
| $a_0$ | - | 11.17 |
| $a_1$ | -0.9231 | -1.553 |
| $a_2$ | -0.4066 | -6.136 |
| $a_3$ | -0.993 | -9.725 |
| $a_4$ | - | 1.85 |
| $b_0$ | -13.96 | -15.11 |
| $b_1$ | 0.3216 | 0.2031 |
| $b_2$ | 0.4532 | 0.5375 |
| $b_3$ | 0.8098 | 0.9144 |
| $b_4$ | - | 0.5953 |
| $c_1$ | 0.002534 | 0.0006091 |
| $c_2$ | 0.003926 | 0.06438 |
| $c_3$ | 0.09924 | 0.0578 |
| $c_4$ | - | 0.001356 |
| $E_{\text{act}}$ | 12047 | 17393 |
| $R_{\text{cor}}$ | 2.7 | 3.0321 |

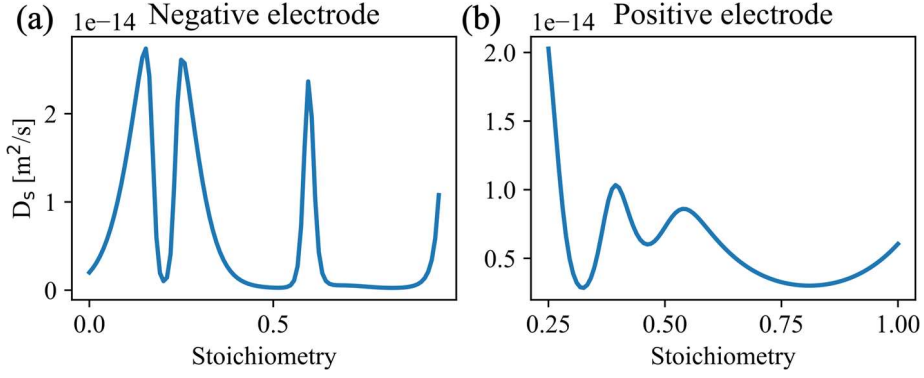

Fig. S2 Solid-phase diffusivity of both electrodes[44].

The electrolyte diffusivity, conductivity, and cation transference number are based on the EC:DMC 1:1 wt% in LiPF$_6$ from Landesfeind and Gasteiger[11]. We manually add the saturation limit of 4M, assuming that at any salt concentration higher than this value, salt precipitation will happen, and the three properties will become constant at higher concentration. The authors need to be careful about the validity of the results higher than that. Coping with the complex precipitation / dissolution dynamics is beyond the scope of this work.

$$c_e^{cor} = \begin{cases} c_e/1000, & c_e < 4000 \\ 4, & c_e \geq 4000 \end{cases} \tag{S7}$$

$$D_e = 10^{-10} \cdot 1470 \cdot e^{1.33 \cdot c_e^{cor}} \cdot e^{-1690/T} \cdot e^{c_e^{cor} \cdot (-563)/T} \tag{S8}$$

$$\kappa_e = 0.1 \cdot 0.798 \cdot (1 + (T - 228)) \cdot c_e^{cor} \cdot \frac{\left(1 - 1.22 \cdot \sqrt{c_e^{cor}} + 0.509 \cdot \left(1 - 4000 \cdot e^{\frac{1000}{T}}\right) \cdot c_e^{cor}\right)}{1 + (c_e^{cor})^4 \cdot \left(0.00379 \cdot e^{\frac{1000}{T}}\right)} \tag{S9}$$

$$t_+^0 = -7.91 + 0.245 \cdot c_e^{cor} + 0.0528 \cdot T + 0.698 \cdot (c_e^{cor})^2 - 0.0108 \cdot c_e^{cor} \cdot T - 8.21 \cdot 10^{-5} \tag{S10}$$
$$\cdot T^2 + 7.43 \cdot 10^{-4} \cdot (c_e^{cor})^3 - 2.22 \cdot 10^{-3} \cdot (c_e^{cor})^2 \cdot T + 3.07 \cdot 10^{-5} \cdot c_e^{cor}$$
$$\cdot T^2$$

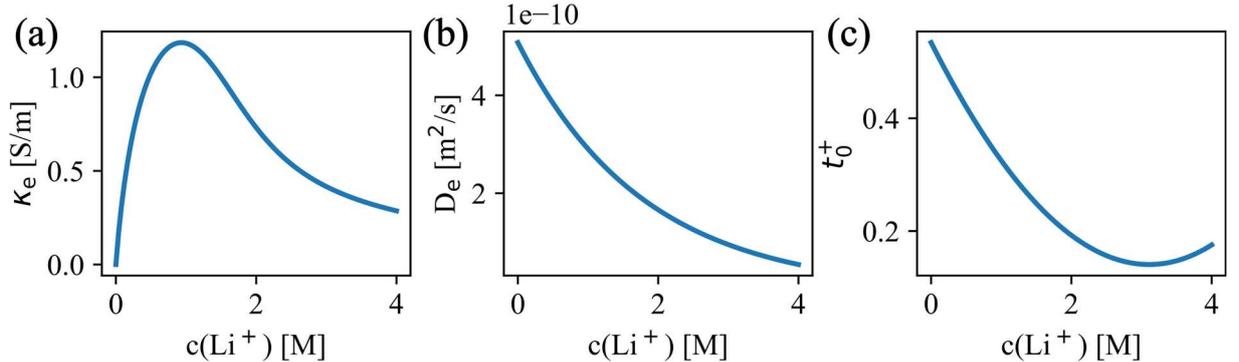

Fig. S3 (a) Electrolyte conductivity, (b) diffusivity, and (c) cation transference number for EC:DMC 1:1 wt% in LiPF$_6$[11].

The measured liquid junction potential is based on EC:EMC in LiPF$_6$ with varying compositions from Jung et al.[23]. In their fitting, the particle fractions ($y_i$) rather than concentrations ($c_i$) are used as basic variables. Notably, the expression of $c_T$ is fitted using the data from Wang et al.[8] at 298.15 K. Based on Eqs. (S11) ~ (S16), the partial derivatives of the $\Delta U$ over $c_{EC}$ and $c_e$ can be obtained and put into the charge conservation equation in the electrolyte. However, we omitted this detailed expression here as it is too long.

$$\Delta U(y_{EC}, y_e) = \left(1 - \frac{y_{EC}}{1 - 2y_e}\right) \Delta U_{1:0}(y_e) + \frac{y_{EC}}{1 - 2y_e} \Delta U_{0:1}(y_e) + \Delta U_{ex}(y_{EC}, y_e) \quad (S11)$$

$$\Delta U_{1:0}(y_e) = \frac{RT}{F}(7.167 - 43.16(y_e)^{0.5} + 185.4(y_e) - 402.4(y_e)^{1.5} + 236.9(y_e)^2 \quad (S12)$$
$$+ 253.7(y_e)^{2.5} - 408.1(y_e)^3 + 2509(y_e)^{3.5} - 2886(y_e)^{4.5}$$
$$+ 1.174 \ln(y_e))$$

$$\Delta U_{0:1}(y_e) = \frac{RT}{F}(3.024 \ln(y_e) + 8.233 - 88.12(y_e) + 477.9(y_e)^2) \quad (S13)$$

$$\Delta U_{ex}(y_{EC}, y_e) = \frac{RT}{F}(y_{EC} - y_{EC}^2 - 2y_e y_{EC})[32.2 - 37.99 y_{EC} - 44.8(1 - y_{EC} - 2y_e)] \quad (S14)$$

$$y_{EC} = \frac{c_{EC}}{c_T}, \quad y_e = \frac{c_e}{c_T}, \quad (S15)$$

$$c_T = 9778 + 1.4631 \cdot c_e + 0.3589 \cdot c_{EC} \quad (S16)$$

The reference total concentration is therefore $\langle c_T \rangle = 13484.224 \, mol/m^3$

For the single-solvent case (normal DFN), the measured liquid junction potential as a function of only salt concentration is:

$$\Delta U(y_e) = \frac{RT}{F}(1.39 \cdot \ln(y_e) + 1.158 - 8.955 \cdot y_e + 164.7 \cdot (y_e)^2) \quad (S17)$$

$$y_e = \frac{c_e}{c_T} \quad (S18)$$

$$c_T = 11130 + 1.379 \cdot c_e \quad (S19)$$

Table S4. The parameters used for the SEI and solvent consumption model.

| Parameter | Unit | Value |
|---|---|---|
| SEI partial molar volume | m³·mol⁻¹ | $9.585 \cdot 10^{-5}$ |
| SEI conductivity ($\sigma_{SEI}$) | S·m⁻¹ | $5 \cdot 10^{-6}$ |
| Initial SEI thickness ($\delta_{SEI}^0$) | m | $2.4724 \cdot 10^{-8}$ |
| Lithium interstitial diffusivity in SEI ($D_{int}$) | m²·s⁻¹ | $5 \cdot 10^{-19}$ |
| Lithium interstitial reference concentration ($c_{int,Li}$) | mol·m⁻³ | 15 |

Note: The product $D_{int} c_{int,Li}$ is the main tuning parameter to control the degradation rate.

## Model validation at 1C GITT and single-solvent cases

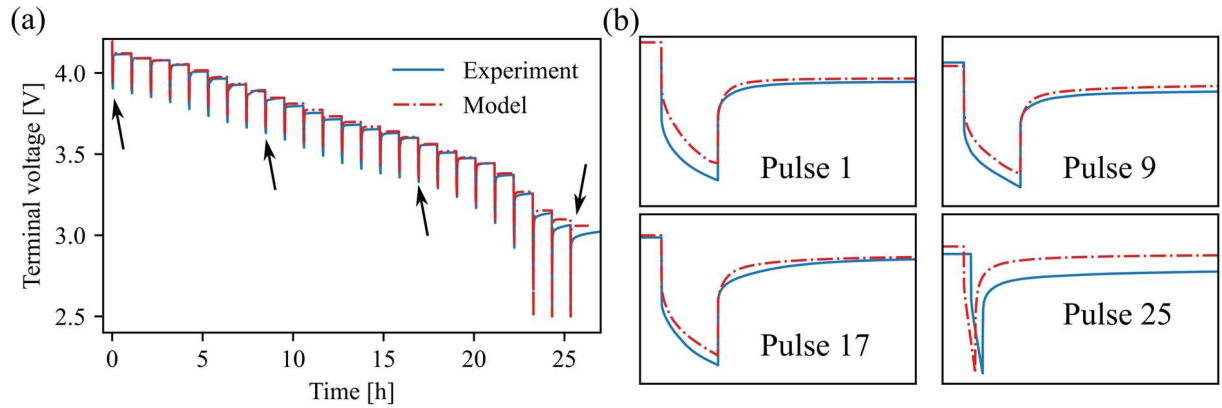

Fig. S4 Validation of the double-solvent model at 1C GITT. (RMSE=29mV)

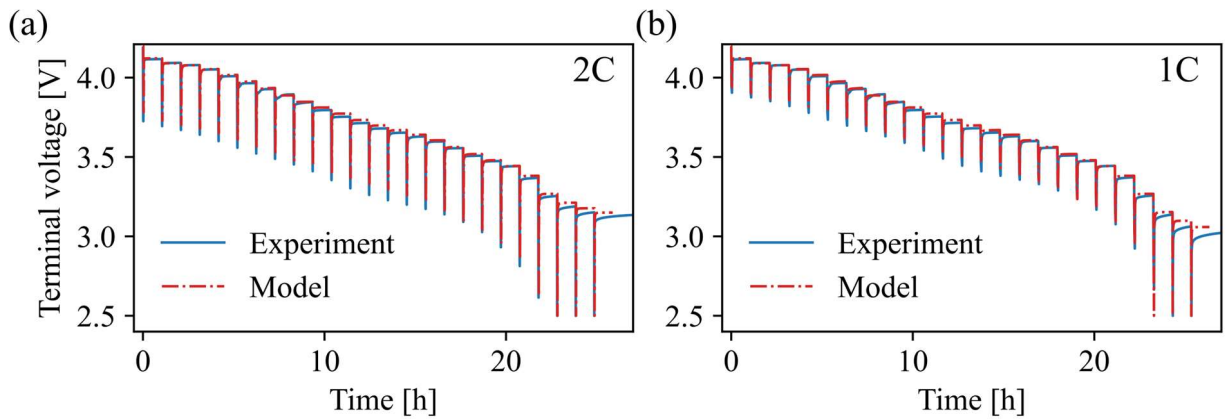

Fig. S5 Validation of the single-solvent model with (a) 2C and 1C GITT. (RMSE are the same with double-solvent cases.)

## Voltage decomposition

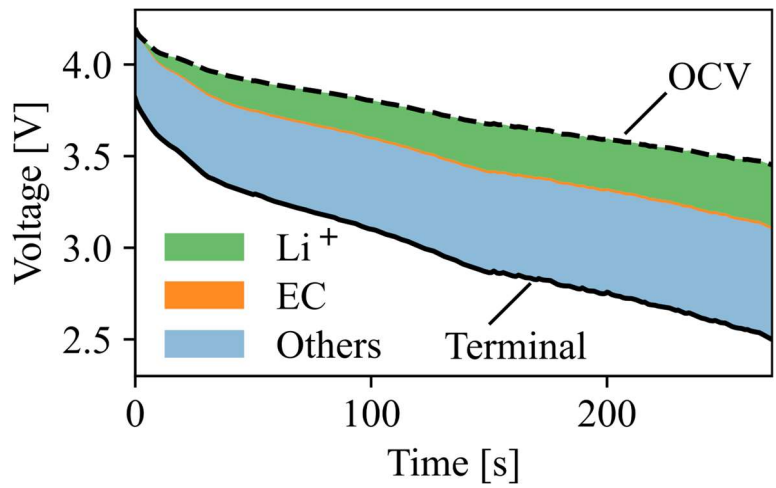

Fig. S6 High $D_\times$, 4.5C discharge, overpotential decomposition.

## Concentration and flux profiles

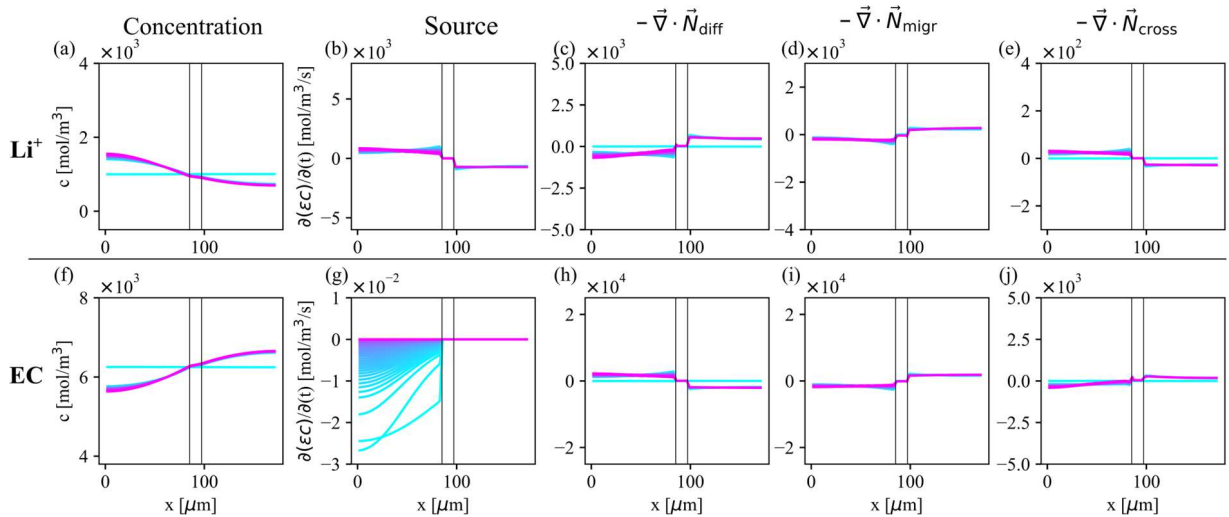

Fig. S7 EC and Li+ concentrations and flux contributions (1C discharge, high $D_\times$)

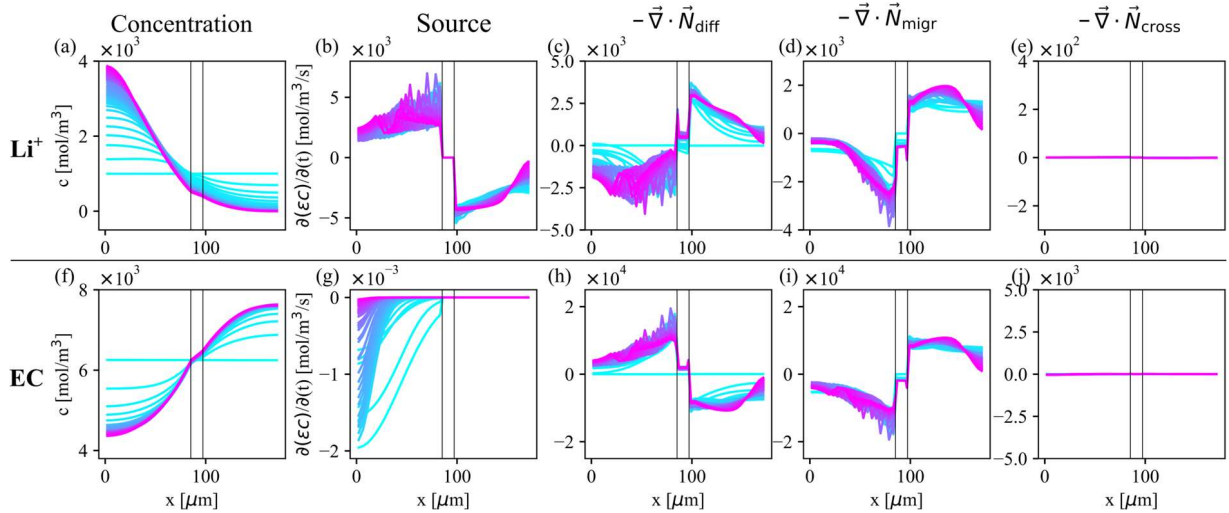

Fig. S8 EC and Li$^+$ concentrations and flux contributions (4.5C discharge, low $D_\times$)

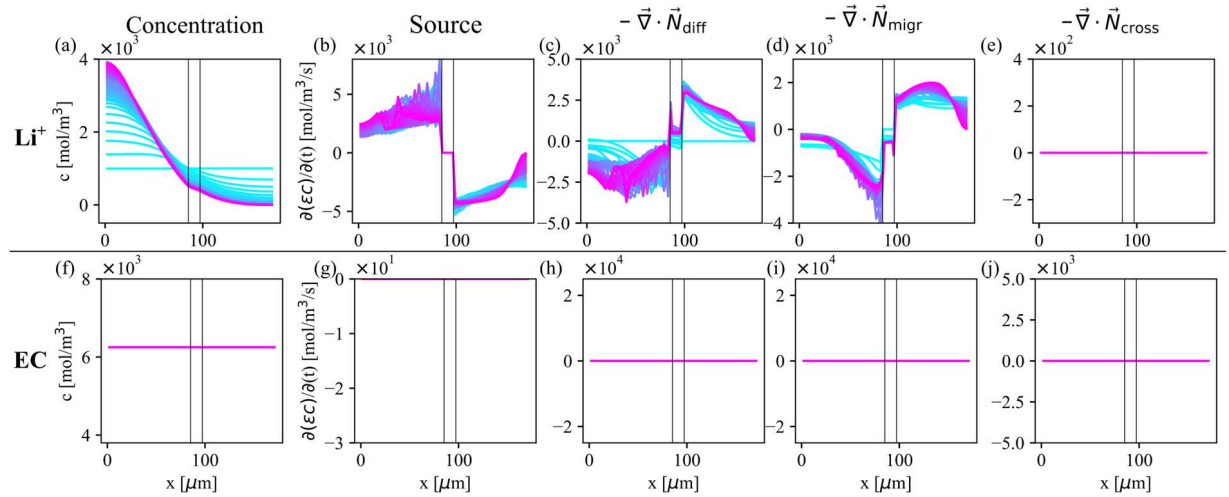

Fig. S9 EC and Li$^+$ concentrations and flux contributions (4.5C discharge, single-solvent case).

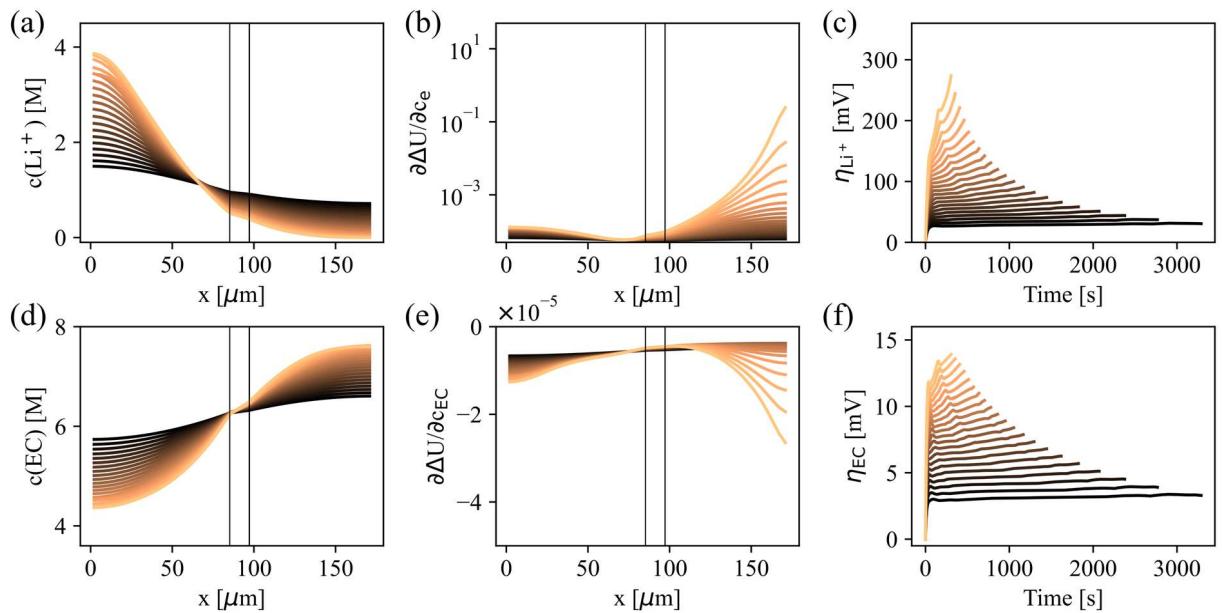

Fig. S10 Low $D_\times$ (different C rates)

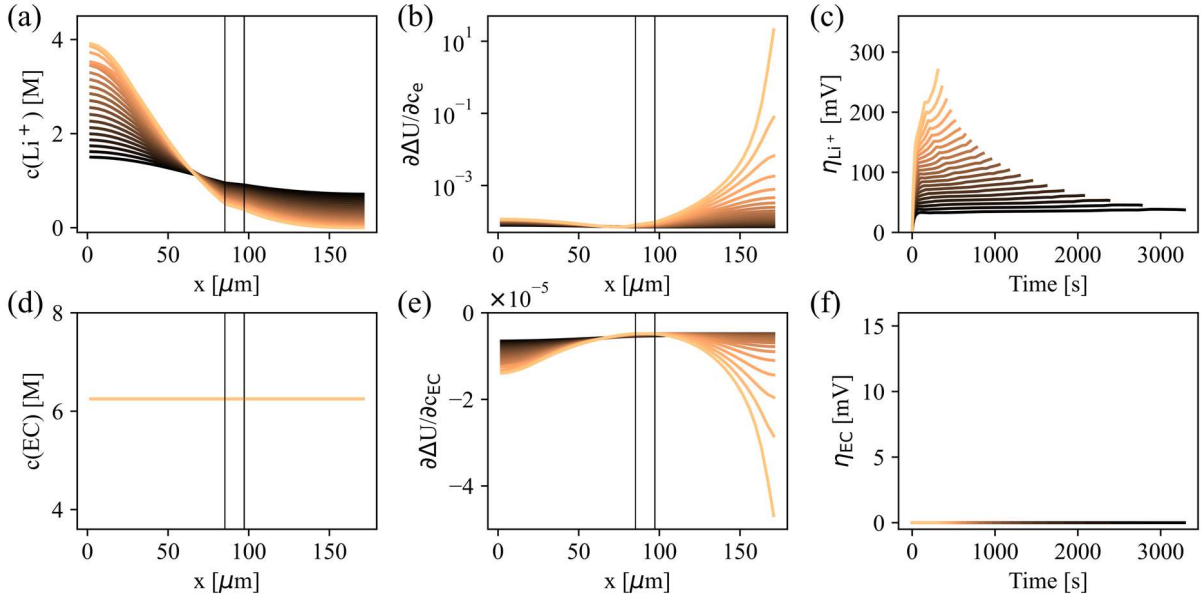

Fig. S11 Single-solvent model (different C rates)

## Parameter sensitivity analysis

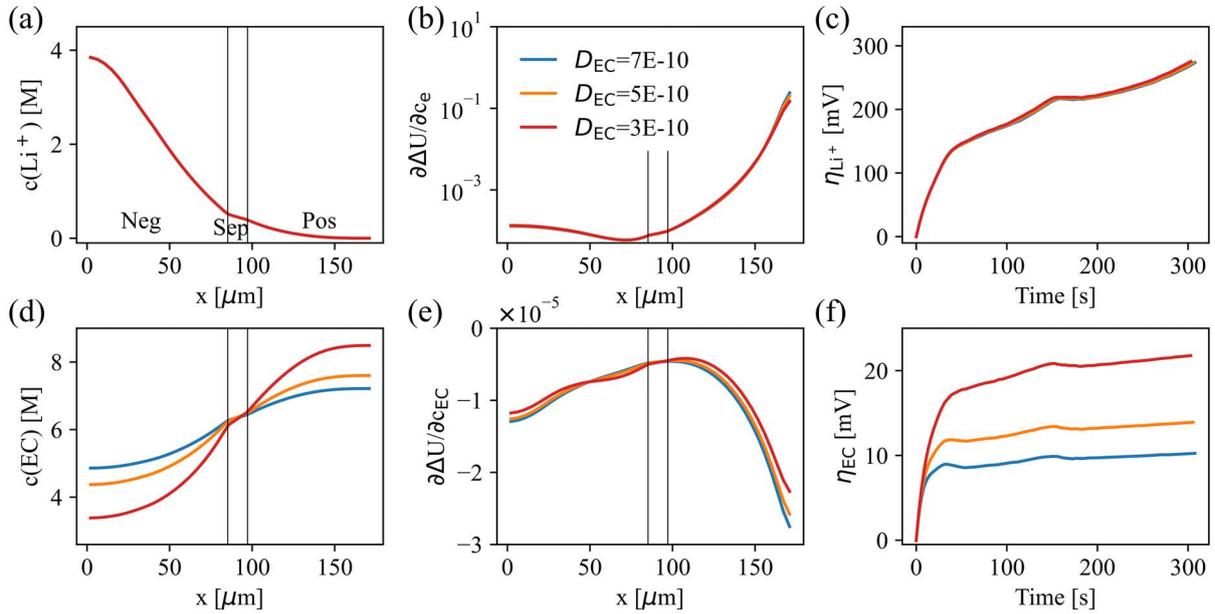

Fig. S12 Concentration profiles and overpotentials for 4.5C discharge. All parameters are the same as the low $D_\times$ case except for $D_{EC}$.

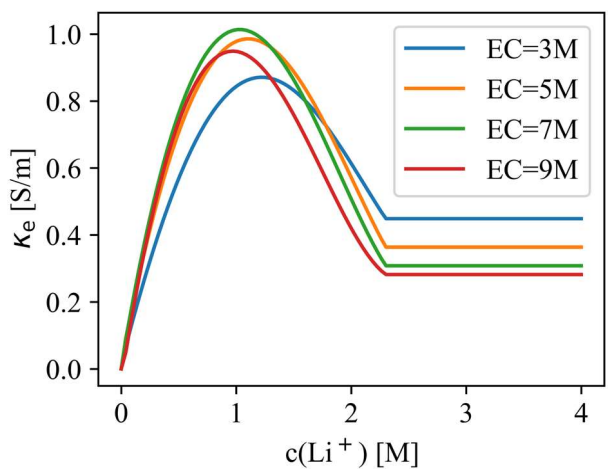

Fig. S13 Electrolyte conductivity of EC:EMC in LiPF6 with varying compositions with constant extrapolation after $c_e \geq 2300$ mol/m$^3$.

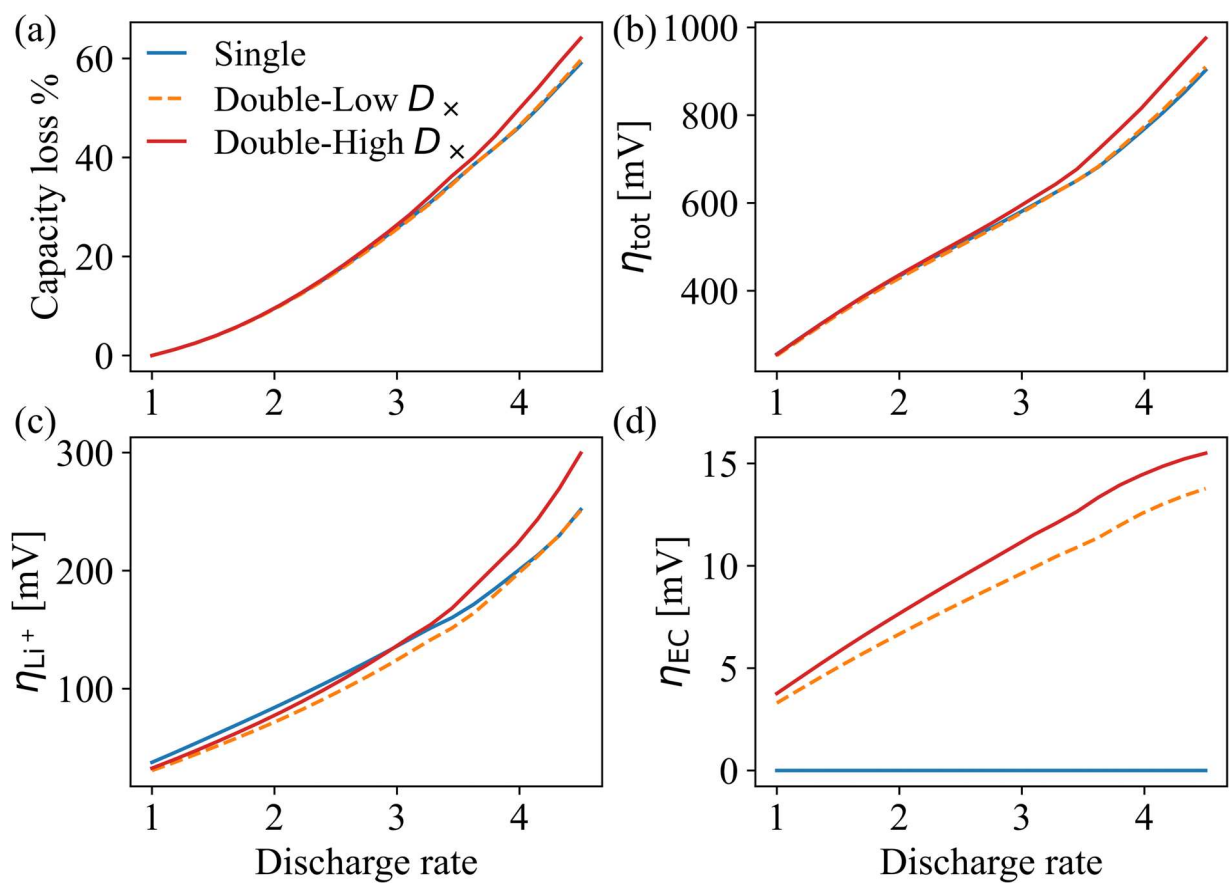

Fig. S14 Rate performance and overpotentials with electrolyte conductivity in Fig. S13.

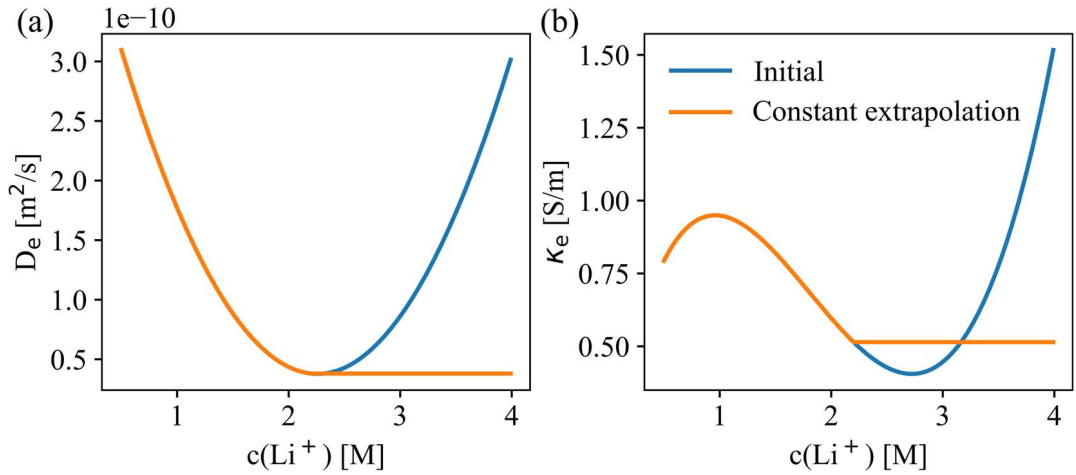

Fig. S15 Electrolyte diffusivity and conductivity from Nyman et al.[35]

**Ageing**

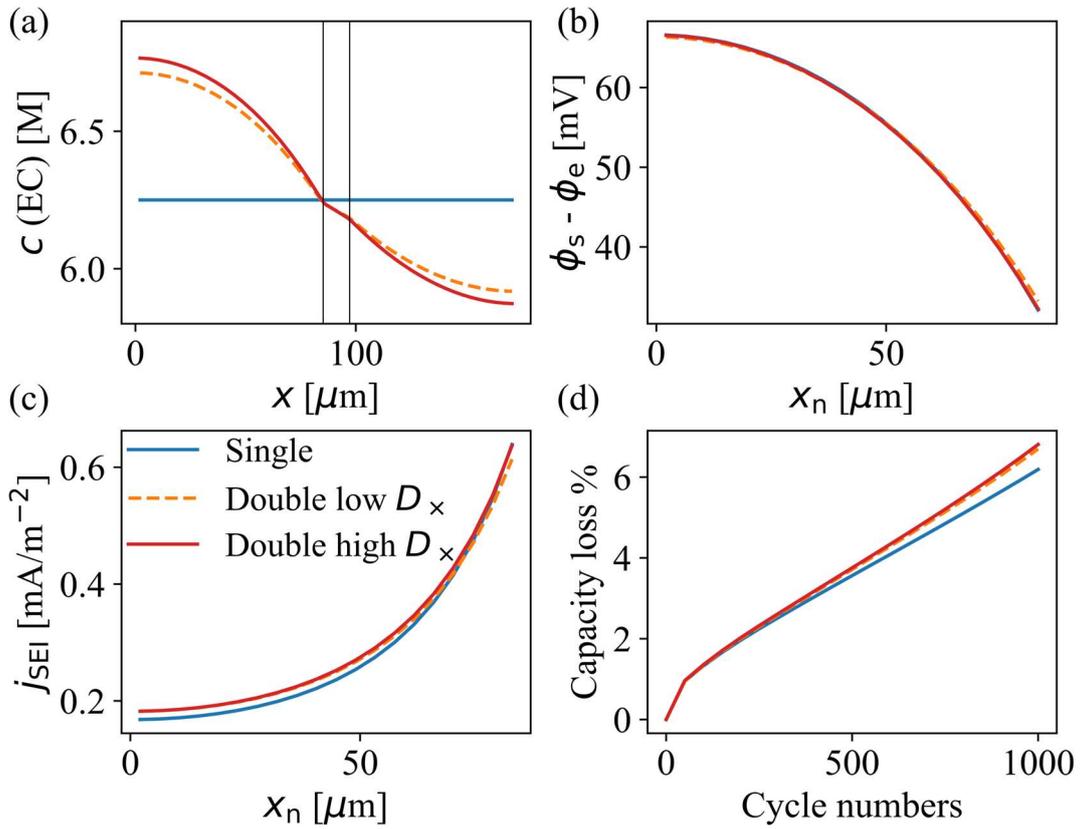

Fig. S16 Model result without solvent consumption

Capacity loss during 1000 cycles for the three cases are 6.19%, 6.70%, 6.81%; difference between single and high $D_\times$ is 0.62%.